\documentclass[aps,prl,reprint,groupedaddressb,twocolumn]{revtex4-2}

\usepackage{graphicx}
\usepackage{color}

\usepackage{bm}
\usepackage{float}
\usepackage{amsthm,amssymb,amsmath,url,braket,amsbsy} 
\usepackage{threeparttable}
\usepackage[utf8]{inputenc}
\usepackage{dirtytalk}
\graphicspath {{./Figure/}}
\usepackage{color}
\usepackage{cancel}

\usepackage{tikz} 
\usetikzlibrary{shapes,arrows,positioning,automata,backgrounds,calc,er,patterns}
\usepackage[compat=1.1.0]{tikz-feynman}

\usepackage{indentfirst} 
\usepackage{bbm}

\usepackage{hyperref}
\hypersetup{
    colorlinks=true,
    linkcolor=blue,
    citecolor=blue,
    urlcolor=blue,
}

\usepackage{units}
\usepackage{comment}
\begin{document}

\title{Circular Photon Drag Effect in Dirac electrons by Quantum Geometry}
\author{Guanxiong Qu}
\affiliation{RIKEN Center for Emergent Matter Science (CEMS), Wako 351-0198, Japan.}

\begin{abstract}
Quantum geometry is a well-established framework for understanding transport and optical responses in quantum materials. In this work, I study the photon drag effect in Dirac electrons using the quantum geometric interpretation of non-vertical optical transitions. Due to the particle-hole symmetry  inherent in Dirac electrons, the shift photon-drag photocurrent is dominated by dissipationless Fermi surface contributions, connected to the dipole of quantum metric tensor. I find that this dipole is significantly enhanced by a small band gap in massive Dirac electrons and remains robust in the massless limit. I demonstrate the existence of a circular shift photon-drag current in the effective Hamiltonian at the $L$-point of bismuth, where the bands exhibit trivial topology, highlighting the ubiquity of the circular photon-drag effect in centrosymmetric materials. 
\end{abstract}


\maketitle

\date{\today}
\section{Introduction}
The geometry of quantum states offers a unified framework for capturing the behavior of electronic systems in response to electromagnetic fields~\cite{RevModPhys.82.1959,ahn_riemannian_2022}. In the static limit, transport responses are described by geometric quantities of a single state, exemplified by the well-known correspondence between  the anomalous Hall effect~\cite{jungwirth_anomalous_2002,onoda_topological_2002} in magnetic materials and the Berry curvature~\cite{berry1984quantal}, which is the imaginary part of quantum geometric tensor~\cite{provost_riemannian_1980}. However, the optical transitions~\cite{aversa_nonlinear_1995,sipe_second-order_2000,hosur_circular_2011,sodemann_quantum_2015,morimoto_semiclassical_2016,ventura_gauge_2017,parker_diagrammatic_2019,ahn_low-frequency_2020,watanabe_chiral_2021} are inherently interband processes, involving a pair of valance and conduction bands. As such, they cannot be adequately described by single-band quantum geometry beyond the two-band approximation; instead, they are captured by constructing Riemannian geometry via identifying interband transition dipole moments as tangent basis vectors~\cite{ahn_riemannian_2022}.

 \begin{figure}[t]
 \includegraphics[width=8cm]{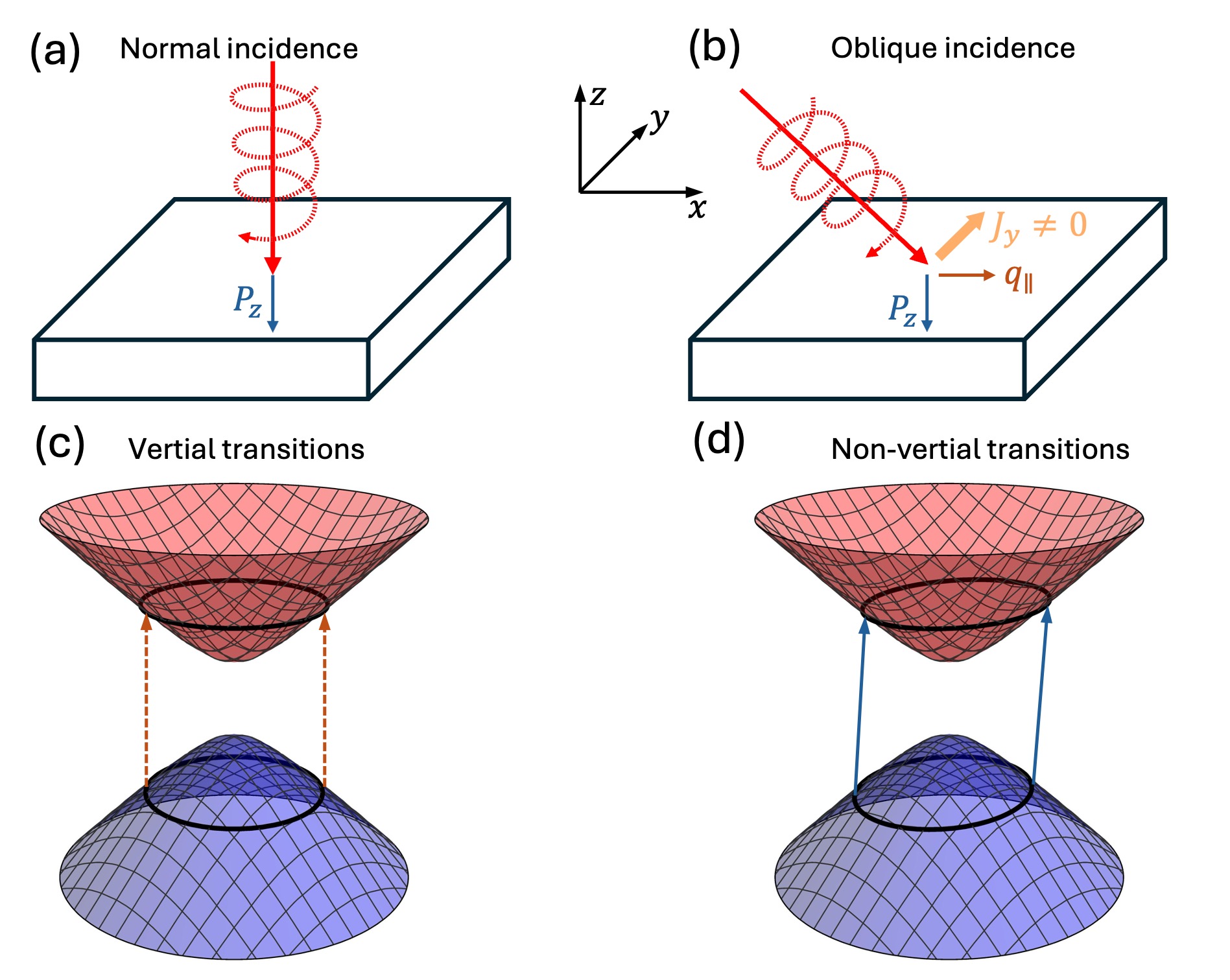}
 \caption{\label{fig:1} 
 Schematic illustrations of photocurrents for (a) normal incidence and (b) oblique incidence of CP light, where $P_z$ denotes the perpendicular component of circular polarization. For normally incident CP light, only (c) vertical optical transitions are induced. In contrast, obliquely incident CP light induces (d) non-vertical optical transitions, transferring in-plane momentum $q_{\parallel}$ from the photon to the electron and generating a transverse charge current. }
 \end{figure}

The bulk photovoltaic effect~\cite{von_baltz_theory_1981,fridkin_bulk_2001,sturman_photovoltaic_2021} renders the need for the heterogeneous structure required in conventional $p$-$n$ junctions, instead relying on intrinsic symmetry breaking within the homogeneous materials. For instance, in non-centrosymmetric materials, which breaks spatial inversion ($\mathcal{P}$) symmetry, the emergent Berry curvature dipole~\cite{sodemann_quantum_2015} and shift vector~\cite{sipe_second-order_2000} enable the generation of  nonlinear injection and shift photocurrents. Notably, the shift vector can be further described by the Levi–Civita connection~\cite{ahn_low-frequency_2020} of the quantum geometric tensor. The intrinsic constraint  of $\mathcal{P}$-symmetry breaking can, however, be relaxed by introducing the photon-drag processes~\cite{ribakovs_theory_1977,goff_theory_2000,entin_theory_2010,shi_geometric_2021}, where  extrinsic momentum transfer from either photons or surface polaritons~\cite{basov_polaritons_2016,kurman_control_2018} facilitates non-vertical optical transitions under oblique light incidence~\cite{hatano_transverse_2009,karch_dynamic_2010,shalygin_circular_2016,strait_revisiting_2019},  as opposed to the vertical optical transitions in conventional bulk photovoltaic effect [Fig.~\ref{fig:1}]. However, the quantum geometric interpretation of photon drag effect (PDE) is still unexplored.

In this article, I present a unified description of PDE in Dirac electrons, based on the Riemannian geometry of optical transitions~\cite{ahn_riemannian_2022}. Owing to the particle-hole symmetry of Dirac electrons, I find that the PDE is dominated by the Fermi surface contribution, featured by a dissipationless shift photon-drag current. In particular, the circular shift current of the PDE~\cite{shalygin_spin_2007,hatano_transverse_2009,karch_dynamic_2010,shalygin_circular_2016,hamara_ultrafast_2023} arises from the product of group velocity and antisymmetric quantum metric connection which shows a dipolar distribution $\bm{k}$-space, [i.e., Eq.~(\ref{eq:4-9})].  The circular shift current remains non-vanishing in trivial band topology that preserves both $\mathcal{P}$- and $\mathcal{T}$-symmetries. I demonstrate  the existence of the circular PDE in bismuth films, where the $L$-electron pockets can be effectively described by the massive Dirac Hamiltonian~\cite{fuseya2015transport,chi_spin_2022}. My findings indicate that the small band gap of massive Dirac electrons is crucial for enhancing the dipolar distribution of antisymmetric quantum metric connection which can be expressed as dipole of quantum metric tensor. These results indicate that the circular shift current of the PDE is universal, irrespective of the band topology, offering a promising probe for the quantum geometry of solids.

\section{Formalism for photon drag effect}
I employ the velocity-gauge approach~\cite{ventura_gauge_2017} to introduce the electric field via minimal coupling with vector potential $\bm{A} (\bm{q}, t)$, accounting for finite photon momentum $\hbar \bm{q}$ in non-vertical transitions~\cite{shi_geometric_2021}:
\begin{align}
\hat{H}_{\text{E}} = -e  A^{\alpha}(\bm{q}, \omega) \hat{v}^\alpha,
\label{eq:2-1}
\end{align}
where $e<0$ is the electron charge and $\hat{v}^\alpha$ is the velocity operator. Summation is implied over repeated indices. The static current response is calculated using the second order density response $\rho^{(2)}_{nm}$ and the velocity matrix $v^\gamma_{nm}$. The photocurrent is usually separated into injection (intraband) and shift (interband) components~\cite{aversa_nonlinear_1995,sipe_second-order_2000}:
\begin{align}
J^{(2),\gamma}_{\text{inj}} = e\sum_{n,\bm{k}} v^{\gamma}_{nn} \rho^{(2)}_{nn}, \quad J^{(2),\gamma}_{\text{shift}}  = e\sum_{\tilde{n} \neq \tilde{m},\bm{k}} v^{\gamma}_{nm} \rho^{(2)}_{mn} ,
\label{eq:2-2}
\end{align}
where $\tilde{n} \neq \tilde{m}$ denotes summation over non-degenerate subspace, i.e., $E_n \neq E_m$.

Through straightforward calculations~\cite{supplementary}, I derive the injection and shift conductivities:
\begin{align}
\label{eq:2-3}
&\sigma^{\gamma;\alpha \beta}_{\text{inj}} (\bm{q}) = \frac{- e^3}{2  \hbar^2 \omega^2 \eta } \sum_{n,m,\bm{k}} f_{nm}(\bm{q})  \notag \\
& \times  \Delta^{\gamma}_{nm}(\bm{q}) v^{\beta}_{nm}(\bm{q}) v^{\alpha}_{mn}(-\bm{q}) \delta(\omega_{nm}(\bm{q}) - \omega), \\
&\sigma^{\gamma; \alpha \beta}_{\text{shift}} (\bm{q}) = \frac{e^3}{2 \hbar^2 \omega^2} \sum_{n,m,\bm{k}} f_{nm} (\bm{q}) \notag \\
& \times \left[  \frac{v^{\beta}_{nm} (\bm{q}) \mathcal{D}_\gamma v^{\alpha}_{mn} (-\bm{q})}{\omega_{nm} (\bm{q}) - \omega - i \eta} +
\frac{ \mathcal{D}_\gamma v^{\beta}_{nm} (\bm{q}) v^{\alpha}_{mn} (-\bm{q})}{\omega_{nm} (\bm{q}) - \omega + i \eta}  \right],
\label{eq:2-4}
\end{align}
where $f_{nm}(\bm{q}) = f_{n,\bm{k} + \bm{q}/2}- f_{m,\bm{k} - \bm{q}/2}  $ represents the difference between Fermi distributions, $\Delta^{\gamma}_{nm}(\bm{q})=v^\gamma_{n,\bm{k}+ \bm{q}/2} -v^\gamma_{m,\bm{k}- \bm{q}/2} $ is the difference in group velocities, and $\omega_{nm}(\bm{q}) = \omega_{n, \bm{k}+ \bm{q}/2 } -\omega_{m, \bm{k}- \bm{q}/2 }$ denotes the band energy difference. $\mathcal{D}_\gamma$ denotes the covariant derivative with degenerate bands~\footnote{The explicit form of covariant derivative on degenerate bands is $[\mathcal{D}_\gamma O ]_{nm}  = \partial_{\gamma} O_{nm} - i \left( \sum\limits_{\tilde{l} = \tilde{n}} A^\gamma_{nl} O_{lm} - \sum\limits_{\tilde{l} = \tilde{m}} O_{nl} A^\gamma_{lm} \right)$ with summation over degenerate subspaces $\tilde{l}=\tilde{n},\tilde{m}$.}. Here, $\omega$ is the photon frequency and $\tau_{ph}=\hbar/\eta$ is phenomenological relaxation time. In the clean limit $\eta \to 0$, the injection and shift currents of PDE depend on $\tau^1_{ph}$ and $\tau^0_{ph}$~\cite{PhysRevB.100.064301}, respectively, with only injection current involving relaxation processes. It should be noted that extrinsic contributions beyond the relaxation time approximation~\cite{du2019disorder} have not been taken into account in the present analysis.

 Considering that the photon wavevector is typically two orders  of magnitude smaller than electron wavevector, Eqs.~(\ref{eq:2-3}\&\ref{eq:2-4}) can be expanded to the first order in the photon wavevector $\bm{q}$:
\begin{align}
\sigma^{\gamma; \alpha \beta} (\bm{q}) = \sigma^{\gamma; \alpha \beta} (0) + q^\tau \chi^{\gamma \tau ;\alpha \beta} + \mathcal{O} (q^2).
\label{eq:2-5}
\end{align}
Here, $ \sigma^{\gamma; \alpha \beta} (0)$ accounts for only vertical transitions, which correspond to linear (LPGE) and circular  (CPGE) photogalvanic effects~\cite{watanabe_chiral_2021}. $\chi^{\gamma \tau ;\alpha \beta}$ represents the first-order PDE, containing solely non-vertical transitions [Fig.~\ref{fig:1} (b)].

\section{Photon drag effect in Dirac Hamiltonian}
Optical responses arise from transitions between valence and conduction bands, wherefore I employ the Dirac Hamiltonian effectively describing such transitions near the band edge:
\begin{align}
H = \Delta \gamma_0 + i  \hbar v_F  k^i \gamma_0 \gamma_i .
\label{eq:3-1}
\end{align}
where $ \gamma_\mu$ are Dirac matrices, $ 2\Delta$ is the band gap, and $v_F$ is the Fermi velocity of the model.

\subsection{Particle-hole symmetry in Dirac Hamiltonian}
The Dirac Hamiltonian possesses charge conjugation symmetry, $\mathcal{C}=i \gamma_2 K$ (where $K$ denotes complex conjugation), which corresponds to the particle-hole symmetry in context of band electrons. The charge conjugation operator transforms a conduction state to a valence state:
\begin{align}
\mathcal{C} u_{c,\sigma} (\bm{k}) =   u_{v,\bar{\sigma}} (\bm{k}) (i \sigma_2)_{ \bar{\sigma} \sigma} ,
\label{eq:3-2}
\end{align}
where $c,v$ denotes the conduction and valance bands, respectively, and $\sigma$ denotes the spin degree of freedom. 

In the $\bm{q}$-expansion of Eqs.~(\ref{eq:2-3}\&\ref{eq:2-4}), both injection and shift currents of the PDE can be divided into \textit{Fermi surface} and \textit{Fermi sea} contributions~\cite{supplementary}. However, due to the particle-hole symmetry of the Dirac Hamiltonian, the Fermi sea contributions to both the injection and shift currents cancel exactly, leaving only the Fermi surface contributions arising from the $\bm{q}$-expansion of the difference in Fermi distributions:
\begin{align}
\chi^{\gamma \tau ;\alpha \beta} \propto \partial_{q^\tau} f_{nm} (\bm{q}) ,
\label{eq:3-3}
\end{align}
where the Fermi surface factor is defined as $d^\tau_{f} \equiv 2/\hbar \partial_{q^\tau} f_{nm} (\bm{q}) = f'_c  v^\tau_c + f'_v v^\tau_v$. It is noteworthy that the \textit{Fermi sea} contribution to the PDE is generally non-vanishing if the particle-hole symmetry between conduction and valance bands is broken.

 \begin{table}[t]
\caption{\label{tab:1} Contributions to the shift photocurrent under on-resonance and off-resonance conditions, along with their corresponding geometric quantities. The detailed formula is presented in \cite{supplementary}.}
\begin{ruledtabular}
\begin{tabular}{lll}
   & $ \text{on-resonance}$ & \text{off-resonance}   \\
  \hline
\text{LP light} & Im$ S^{+,\beta \alpha;\gamma}_{\tilde{c} ; \tilde{v}}$ & Re$S^{+,\beta \alpha;\gamma}_{\tilde{c} ; \tilde{v}}$,  $g^{\beta \alpha}_{\tilde{c} ; \tilde{v}}$ \\
\text{CP light} & Re$S^{-,\beta \alpha;\gamma}_{\tilde{c} ; \tilde{v}}$ & Im$S^{-,\beta \alpha;\gamma}_{\tilde{c} ; \tilde{v}}$, $\Omega^{\beta \alpha}_{\tilde{c} ; \tilde{v}}$  \\
\end{tabular}
\end{ruledtabular}
\end{table}
 \begin{figure}[t]
 \includegraphics[width=8cm]{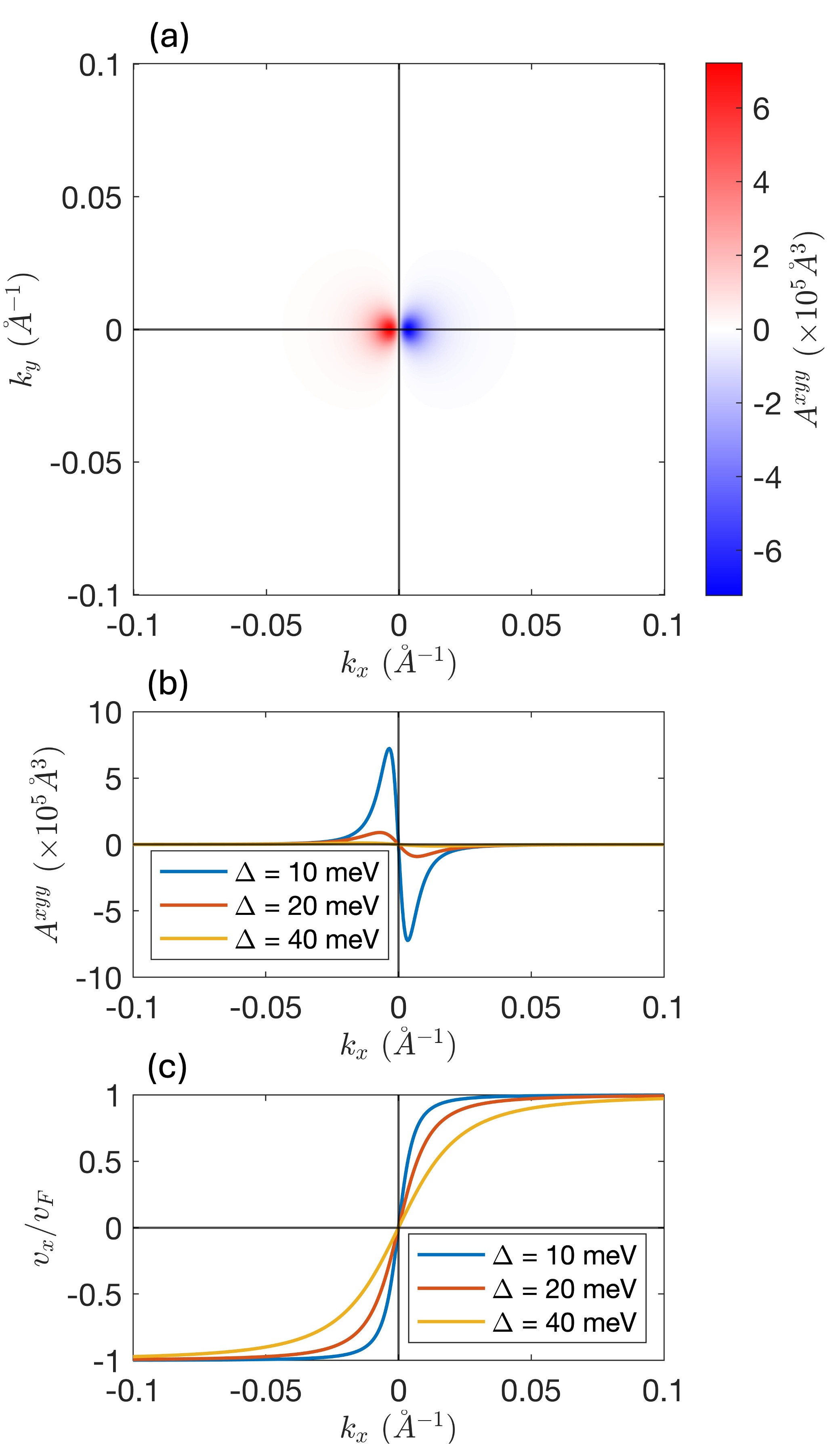}
 \caption{\label{fig:2} (a) Reciprocal space mapping of the antisymmetric connection \( S^{-,xyy} \) in the \( k_z = 0 \) surface. (b) antisymmetric connection \( S^{-,xyy} \) and (c) group velocity \( v_{x,c} \) along the high symmetry line defined by \( k_y = k_z = 0 \) for various gap sizes. The Fermi velocity \( v_F \) is set to \( 2.5 \times 10^5 \) m$\cdot$s$^{-1}$.
 }
 \end{figure}
 \subsection{Injection current of photon drag effect}
 The injection current of the PDE in the Dirac Hamiltonian reads
\begin{align}
\chi^{\gamma \tau ;\alpha \beta}_{\text{inj.}} 
&=   -\frac{ \pi e^3 \tau_{ph} }{  4 \hbar^2 \omega^2  } \sum_{ \bm{k} } d^\tau_{f}  \omega_{cv}^2 Q^{\beta \alpha}_{\tilde{c} ; \tilde{v}}  \Delta^{\gamma}_{cv}   \delta (  \omega_{cv} -  \omega  ) ,
\label{eq:4-1}
\end{align}
where  a compact form can be obtained by employing the quantum geometric tensor~\cite{provost_riemannian_1980}:
\begin{align}
Q^{\beta \alpha}_{\tilde{c} ; \tilde{v}} = \sum_{ \sigma, \tau}  A^{\beta}_{c,  \sigma ; v, \tau}   A^{\alpha}_{v,\tau;  c,\sigma}. 
\label{eq:4-2}
\end{align}
Here, $A^{\beta}_{c, \sigma ; v, \tau}$ is the interband Berry connection outside the degenerate subspace, which equals to interband matrix element of position operator~\cite{blount1962formalisms} and serves as a tangent basis vector for constructing complex Riemannian structure of optical transitions~\cite{ahn_riemannian_2022}. Note that it is directly related to the interband matrix elements of velocity operator, $v^{\beta}_{c, \sigma ; v, \tau}= i \omega_{cv} A^{\beta}_{c, \sigma ; v, \tau}$.

Second-order optical responses can be classified as linearly polarized (LP) and circularly polarized (CP)  components through symmetrization and anti-symmetrization of electric field orientation (i.e., $\alpha,\beta$ in our notation). Consequently, the linear injection photocurrent and circular injection photocurrent are directly connected with the quantum metric and interband Berry curvature:
\begin{align}
Q^{\beta \alpha}_{\tilde{c} ; \tilde{v}} 
= g^{\beta \alpha}_{\tilde{c} ; \tilde{v}} -  \frac{i}{ 2 } \Omega^{\beta \alpha}_{\tilde{c} ; \tilde{v}}.
\label{eq:4-3}
\end{align}
The circular injection current of the PDE  is given by 
\begin{align}
\chi^{\gamma \tau ; \sigma }_{C,\text{inj.}} 
&= -  \frac{   \epsilon_{\alpha \beta \sigma} \pi e^3  \tau_{ph} }{ 8  \hbar^2 \omega^2  } \sum_{ \bm{k} } d^\tau_{f}  \omega_{cv}^2 \Delta^{\gamma}_{cv}  \Omega^{\beta \alpha}_{\tilde{c} ; \tilde{v}}   \delta (  \omega_{cv} -  \omega  ).
\label{eq:4-4}
\end{align}
However, in the Dirac Hamiltonian, which preserves both time  and spatial inversion symmetries, the interband Berry curvature vanishes exactly in whole $\bm{k}$-space ($\Omega^{\beta \alpha}_{\tilde{c} ; \tilde{v}} (\bm{k}) = 0$), leading to vanish of the circular injection current. Note that a non-zero circular injection current of  PDE can be achieved by breaking either of the $\mathcal{T}$- or $\mathcal{P}$-symmetries. In particular,  non-centrosymmetric materials~\cite{xu_electrically_2018} or heterostructures~\cite{duan_berry_2023} that have dipolar distributions of interband Berry curvature $ \Omega^{\beta \alpha}_{\tilde{c} ; \tilde{v}}$ can activate the circular injection photon-drag current.

\subsection{Shift current of photon drag effect}
The shift current of the PDE in the Dirac Hamiltonian can also be expressed in a  compact form:
\begin{align}
&\chi^{\gamma \tau ; \alpha \beta}_{\text{shift}}  = \frac{e^3  }{ 4 \hbar \omega^2}   \sum_{\bm{k}}  \omega^2_{cv}  d^\tau_{f} \notag \\
&\times \left[ \frac{ C^{\beta \gamma \alpha}_{\tilde{c},\tilde{v}} + \frac{\Delta^\gamma_{cv} }{\omega_{cv}} Q^{\beta \alpha}_{\tilde{c} ; \tilde{v}} }{ \omega_{cv} -\omega - i \eta} +  \frac{  ( C^{\alpha \gamma \beta}_{\tilde{c},\tilde{v}} )^*  + \frac{\Delta^\gamma_{cv} }{\omega_{cv}} ( Q^{ \alpha \beta}_{\tilde{c} ; \tilde{v}} )^*  }{ \omega_{cv} -\omega + i \eta }  \right],
\label{eq:4-5}
\end{align}
where $C^{\beta \gamma \alpha}_{\tilde{c},\tilde{v}}$ is the quantum metric connection~\cite{nakahara_geometry_2003,ahn_riemannian_2022}, defined as
\begin{align}
C^{\beta \gamma \alpha}_{\tilde{c},\tilde{v}} &= \sum_{\sigma,\tau}  A^{\beta}_{c,\sigma; v, \tau} \mathcal{D}_\gamma  A^{\alpha}_{ v, \tau;c,\sigma} . 
\label{eq:4-6}
\end{align}
 Note that $C^{\beta \gamma \alpha}_{\tilde{c},\tilde{v}}$ is torsionless (Levi–Civita connection~\cite{nakahara_geometry_2003}), since the two-band Dirac model does not contain three-band virtual transitions~\cite{ahn_riemannian_2022}.

The LP and CP responses are associated with the symmetrization or anti-symmetrization of quantum geometric connection,
\begin{align}
\label{eq:4-7}
S^{\pm,\beta \gamma \alpha}_{\tilde{c},\tilde{v}} &= C^{\beta \gamma \alpha}_{\tilde{c},\tilde{v}} \pm C^{\alpha \gamma \beta}_{\tilde{c},\tilde{v}}, 
\end{align}
which are referred to as symmetric and antisymmetric connections, respectively, in the following sections.

\onecolumngrid
\begin{center}
 \begin{figure*}
 \centering
 \includegraphics[width=\textwidth]{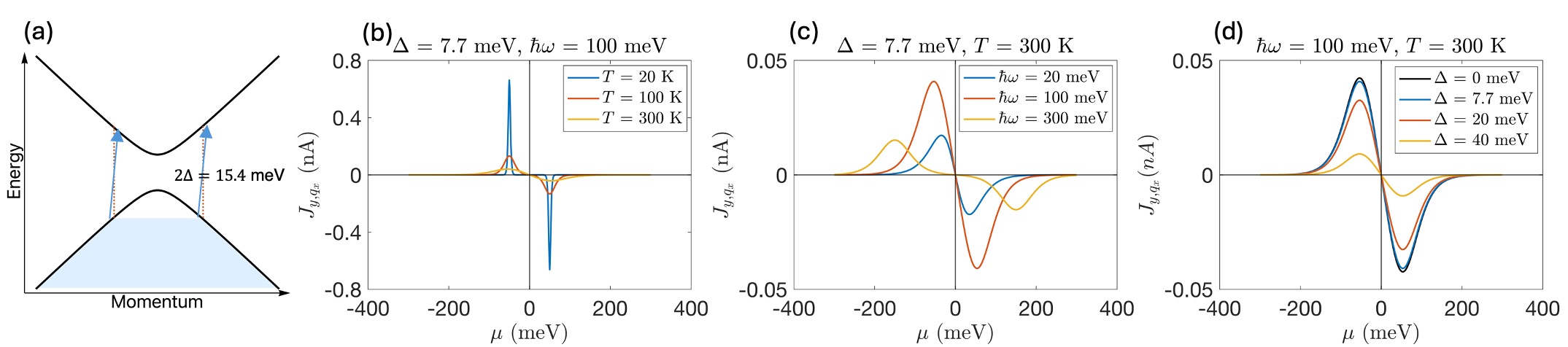}
 \caption{\label{fig:3} (a) Schematic plot of the band dispersion for massive Dirac electrons and non-vertical optical transitions (solid arrow lines). (b-d) Shift current from the circular PDE, $ J_{y,q_x} $, as a function of chemical potential \( \mu \), plotted for (b) various temperatures $ T $, (c) different photon frequencies \( \omega \), and (d) a range of gap sizes \( 2\Delta \). $J_{y,q_x}$ denotes charge current along y-axis with momentum transfer along x-axis. The calculation is adapted for \( L \)-electrons in bismuth, with \( v_F = 2.5 \times 10^5 \) m$\cdot$s$^{-1}$ and \( \Delta = 7.7 \) meV~\cite{chi_spin_2022}. For the experimental setup, I use a 405 nm laser at a \( 45^\circ \) oblique incidence, illuminating bismuth wires with a cross-sectional area of \( A = 0.5 \) mm $\times$ 30 nm. The electric field strength is \( |E| = 1.9 \times 10^3/\sqrt{2} \) V$\cdot$m$^{-1}$, and the in-plane momentum transfer is \( |q_{\parallel}| = 2\pi/(\sqrt{2} \times 405)$ nm$^{-1}$.
}
 \end{figure*}
\end{center}
\twocolumngrid
 
Using these geometric quantities, the shift current of the PDE can be classified into LP/CP responses under on-/off-resonance conditions, as summarized in Tab.\ref{tab:1}. Notably, since the vanish of interband Berry curvature in the Dirac Hamiltonian, its quantum geometric connection has no symplectic form~\cite{ahn_riemannian_2022}, meaning that $C^{\beta \gamma \alpha}_{\tilde{c},\tilde{v}} $ is pure real.  Consequently, I focus on the shift current of CP-light response under the resonance condition ($\omega_{cv} \simeq \omega$):
\begin{align}
\chi^{\gamma\tau; \sigma}_{C,\text{shift}}  
&= - \frac{   \epsilon_{\alpha \beta \sigma}  \pi e^3   }{ 4 \hbar \omega^2}   \sum_{\bm{k}}  \omega^2_{cv} d^\tau_{f,cv} \text{Re}   \left(S^{-,\beta \gamma \alpha}_{{\tilde{c},\tilde{v}}} \right)  \delta (\omega_{cv} -\omega)  ,
\label{eq:4-9}
\end{align}
which integrates the product of group velocity (embedded in $ d^\tau_{f,cv}$) and antisymmetric connection $S^{-,\beta \gamma \alpha}_{{\tilde{c},\tilde{v}}} $ over the entire $\bm{k}$-space.

\subsection{Antisymmetric connection in Dirac Hamiltonian}
The antisymmetric connection $S^{-,\beta \gamma \alpha}_{{\tilde{c},\tilde{v}}}$ in the Dirac Hamiltonian is subject to constraints imposed by its  $\mathcal{T}$ and $\mathcal{P}$ symmetries~\cite{supplementary}, expressed as
\begin{align}
S^{-,\beta \gamma \alpha}_{{\tilde{c},\tilde{v}}} (\bm{k}) =  - S^{-,\beta \gamma \alpha}_{{\tilde{c},\tilde{v}}} (-\bm{k}) =   - S^{-,\beta \gamma \alpha}_{\tilde{v},\tilde{c}} (-\bm{k}) .
\label{eq:5-1}
\end{align}
However, $S^{-,\beta \gamma \alpha}_{{\tilde{c},\tilde{v}}}$ does not vanishes at every $\bm{k}$ point, whose explicit form is given by 
\begin{align}
S^{-,\beta \gamma \alpha}_{{\tilde{c},\tilde{v}}} =   \frac{v_F^4}{2 \omega_k^4} \left(   \delta_{\gamma \alpha}  k^\beta -   \delta_{\gamma \beta}  k^\alpha \right),
\label{eq:5-2}
\end{align}
where $ \omega_k = \sqrt{v_F^2k^2 +\tilde{\Delta}^2}$ is the band dispersion of the Dirac Hamiltonian with the energy gap $\tilde{\Delta} \equiv \Delta/\hbar$  normalized by the reduced Planck constant.

Figure~\ref{fig:2} (a) shows the reciprocal space mapping of $S^{-,x y y}$ in the $k_z=0$ plane. Due to the Dirac Hamiltonian preserves $\mathcal{P}$-inversion symmetry, the antisymmetric connection is odd with respect to $\bm{k}$ and exhibits a dipolar distribution. The dipolar peaks of antisymmetric connection are significantly enhanced by the small gap $2\Delta$ in the massive Dirac electrons [Fig.~\ref{fig:2} (b)]. The magnitude of dipolar peak in the antisymmetric connection  is inversely proportional to cube of gap size, following the relation $|S_{\text{max}}^{-,\beta \gamma \alpha}| \propto (\hbar v_F/\Delta)^{3}$. For massive Dirac fermions, a small gap size of approximately $\sim 20$ meV can result in dipolar peaks of antisymmetric connection as large as $10^5$ $\AA^3$. Furthermore, the group velocity embedded in $d^\tau_{f}$ is also a  odd function with respect to $\bm{k}$ [Fig.~\ref{fig:2} (c)]. Consequently, the integration of the product of group velocity and antisymmetric connection over $\bm{k}$-space yields a non-zero shift photon-drag current with CP light. In contrast, the CPGE vanishes in centrosymmetric materials, which is direct consequence of the antisymmetric connection being odd with respect to $\mathcal{P}$-inversion [Eq.~(\ref{eq:5-1})].

\subsection{Dipole of quantum metric tensor}

For a trivial band structure, where the interband Berry curvature vanishes everywhere in reciprocal space and quantum geometric connection has no symplectic form, the quantum metric connection can be expressed as the Chrisoffel symbol of the first kind~\cite{nakahara_geometry_2003,ahn_low-frequency_2020}:
\begin{align}
C_{\beta \gamma \alpha} = \Gamma_{\beta \gamma \alpha} \equiv \frac{1}{2} \left( \partial_\gamma g_{\beta \alpha} + \partial_{\alpha} g_{\beta \gamma} -  \partial_{\beta} g_{\alpha \gamma}  \right),
\label{eq:5-3}
\end{align}
where the band indices are omitted, and the covariant indices are lowered for clarity. Consequently, the symmetric and antisymmetric connections can be expressed as dipole of quantum metric tensor $g_{\beta \alpha}$~\cite{gao_nonreciprocal_2019,das_intrinsic_2023,kaplan_unification_2024}:
\begin{align}
S^{+}_{\beta \gamma \alpha} =  \partial_\gamma g_{\beta \alpha}, \quad S^{-}_{\beta \gamma \alpha} =  \partial_{\alpha} g_{\beta \gamma} - \partial_{\beta} g_{\alpha \gamma}.
\label{eq:5-4}
\end{align}
It is important to note that $S^{\pm}_{\beta \gamma \alpha}$ are not naturally zero, even in trivial band topology.  For the symmetric connection $ S^{+,\beta}_{\gamma\alpha}$, it is evident that  $ g_{\beta\alpha}$ is not trivially zero throughout the entire $\bm{k}$-space. However, the on-resonant linear shift current of the PDE requires the symplectic form of the quantum metric connection, $\text{Im}S^{+,\beta}_{\gamma\alpha}$, necessitating non-trivial band topology to provide non-zero Berry curvature, e.g., by breaking the $\mathcal{T}$-inversion symmetry~\cite{shi_geometric_2021}. In contrast, the constraints on the on-resonant shift current of circular PDE are less stringent, as $ \text{Re}S^{-,\beta}_{\gamma\alpha} $ can be finite even in trivial band topology. Therefore, the circular shift current of the PDE can be considered as a ubiquitous optical property.

\section{$L$-electron in bismuth}
In this section, I employ the Dirac Hamiltonian as effective model to describe the $L$-electron ellipsoids in bismuth~\cite{fuseya2015transport,chi_spin_2022} and estimate the circular shift current of PDE. 
Bismuth is a prototypical Dirac semimetal with trivial band topology, featured by three electron pockets located at the $L$ points and a gap size of approximately 15.4 meV~\cite{PhysRevB.10.771}. The transport properties in bismuth are dominated by the $L$-electron ellipsoids, due to their small effective mass~\cite{PhysRevB.84.115137}.

Figure~\ref{fig:3} presents the circular shift current as function of chemical potential $\mu$. The circular shift current is odd function of $\mu$, due to the opposite group velocities between conduction and valance bands $\bm{v}_c(\bm{k}) = - \bm{v}_v(\bm{k})$. The temperature dependence arises from the derivative of the Fermi distribution, $d^\tau_{f}$, where temperature acts as  a smearing factor diminishes the circular shift  current at high temperature [Fig.~\ref{fig:3} (b)]. Under the resonance condition $\omega_{cv}=\omega$ and with the restriction imposed by $d^\tau_{f}$, the circular shift current peaks at chemical potential $2 |\mu| = \hbar \omega$ [Fig.~\ref{fig:3} (c)]. Moreover, the peak value of circular shift current is not monotonic with photon energy $\hbar\omega$; instead, it reaches its maximum near $\hbar\omega \sim 4 \Delta$. Figure~\ref{fig:3} (d) shows the dependence of the circular shift current on the gap size while keeping Fermi velocity fixed. It is confirmed that the circular shift current decreases with increasing the gap size,  reflecting the reduction in the antisymmetric connection $S^{-,xyy}$ [Fig.~\ref{fig:2} (b)]. Notably, the circular shift current does not vanish for massless Dirac electrons (i.e., $\Delta=0$). The asymmetric connection does not vanish with massless Dirac electrons but exhibits a singular form at Dirac point ($\bm{k}=0$). However, for finite photon frequency $\omega \neq 0$, the energy-momentum-conversing contours [Fig.~\ref{fig:3} (a)] lies outside the Dirac point where $S^{-,xyy}$ remains finite, resulting in a non-zero circular shift photon-drag current in massless Dirac electrons.

\section{Conclusion}
I have shown that the PDE can be described within a unified framework of quantum geometry of optical transitions. In Dirac electrons, the non-vertical optical transitions of the PDE are predominantly located at Fermi surface due to the presence of particle-hole symmetry. Notably, in such $\mathcal{P}$-symmetric systems, the shift current does not vanish because both the antisymmetric quantum metric connection $S^{-}_{\beta \gamma \alpha}$ and group velocity are odd with respect to $\mathcal{P}$-inversion, where the dipolar distribution of quantum metric connection is connected with the quantum geometric tensor dipole. The ubiquity of quantum geometric tensor dipole suggests the universality of circular PDE even in bands with trivial topology.

\textit{Note added.}Recently, I became aware of an independent study by Ying-Ming Xie and Naoto Nagaosa related to this work [44]. They exemplify the photon drag effect in two-dimensional topological insulators and magnetic Weyl semimetals. In my paper, I emphasize the existence of circular PDE in bands with trivial topology, arising from the dipole structure of the quantum metric tensor.

\begin{acknowledgments}
The author thanks Gen Tatara, Masamitsu Hayashi, and Daniel Loss for their insightful discussions.

\end{acknowledgments}


\bibliography{PDE_in_Dirac}

\begin{thebibliography}{45}%
\makeatletter
\providecommand \@ifxundefined [1]{%
 \@ifx{#1\undefined}
}%
\providecommand \@ifnum [1]{%
 \ifnum #1\expandafter \@firstoftwo
 \else \expandafter \@secondoftwo
 \fi
}%
\providecommand \@ifx [1]{%
 \ifx #1\expandafter \@firstoftwo
 \else \expandafter \@secondoftwo
 \fi
}%
\providecommand \natexlab [1]{#1}%
\providecommand \enquote  [1]{``#1''}%
\providecommand \bibnamefont  [1]{#1}%
\providecommand \bibfnamefont [1]{#1}%
\providecommand \citenamefont [1]{#1}%
\providecommand \href@noop [0]{\@secondoftwo}%
\providecommand \href [0]{\begingroup \@sanitize@url \@href}%
\providecommand \@href[1]{\@@startlink{#1}\@@href}%
\providecommand \@@href[1]{\endgroup#1\@@endlink}%
\providecommand \@sanitize@url [0]{\catcode `\\12\catcode `\$12\catcode
  `\&12\catcode `\#12\catcode `\^12\catcode `\_12\catcode `\%12\relax}%
\providecommand \@@startlink[1]{}%
\providecommand \@@endlink[0]{}%
\providecommand \url  [0]{\begingroup\@sanitize@url \@url }%
\providecommand \@url [1]{\endgroup\@href {#1}{\urlprefix }}%
\providecommand \urlprefix  [0]{URL }%
\providecommand \Eprint [0]{\href }%
\providecommand \doibase [0]{https://doi.org/}%
\providecommand \selectlanguage [0]{\@gobble}%
\providecommand \bibinfo  [0]{\@secondoftwo}%
\providecommand \bibfield  [0]{\@secondoftwo}%
\providecommand \translation [1]{[#1]}%
\providecommand \BibitemOpen [0]{}%
\providecommand \bibitemStop [0]{}%
\providecommand \bibitemNoStop [0]{.\EOS\space}%
\providecommand \EOS [0]{\spacefactor3000\relax}%
\providecommand \BibitemShut  [1]{\csname bibitem#1\endcsname}%
\let\auto@bib@innerbib\@empty
\bibitem [{\citenamefont {Xiao}\ \emph {et~al.}(2010)\citenamefont {Xiao},
  \citenamefont {Chang},\ and\ \citenamefont {Niu}}]{RevModPhys.82.1959}%
  \BibitemOpen
  \bibfield  {author} {\bibinfo {author} {\bibfnamefont {D.}~\bibnamefont
  {Xiao}}, \bibinfo {author} {\bibfnamefont {M.-C.}\ \bibnamefont {Chang}},\
  and\ \bibinfo {author} {\bibfnamefont {Q.}~\bibnamefont {Niu}},\ }\bibfield
  {title} {\bibinfo {title} {Berry phase effects on electronic properties},\
  }\href {https://doi.org/10.1103/RevModPhys.82.1959} {\bibfield  {journal}
  {\bibinfo  {journal} {Rev. Mod. Phys.}\ }\textbf {\bibinfo {volume} {82}},\
  \bibinfo {pages} {1959} (\bibinfo {year} {2010})}\BibitemShut {NoStop}%
\bibitem [{\citenamefont {Ahn}\ \emph {et~al.}(2022)\citenamefont {Ahn},
  \citenamefont {Guo}, \citenamefont {Nagaosa},\ and\ \citenamefont
  {Vishwanath}}]{ahn_riemannian_2022}%
  \BibitemOpen
  \bibfield  {author} {\bibinfo {author} {\bibfnamefont {J.}~\bibnamefont
  {Ahn}}, \bibinfo {author} {\bibfnamefont {G.-Y.}\ \bibnamefont {Guo}},
  \bibinfo {author} {\bibfnamefont {N.}~\bibnamefont {Nagaosa}},\ and\ \bibinfo
  {author} {\bibfnamefont {A.}~\bibnamefont {Vishwanath}},\ }\bibfield  {title}
  {\bibinfo {title} {Riemannian geometry of resonant optical responses},\
  }\href {https://doi.org/10.1038/s41567-021-01465-z} {\bibfield  {journal}
  {\bibinfo  {journal} {Nat. Phys.}\ }\textbf {\bibinfo {volume} {18}},\
  \bibinfo {pages} {290} (\bibinfo {year} {2022})}\BibitemShut {NoStop}%
\bibitem [{\citenamefont {Jungwirth}\ \emph {et~al.}(2002)\citenamefont
  {Jungwirth}, \citenamefont {Niu},\ and\ \citenamefont
  {MacDonald}}]{jungwirth_anomalous_2002}%
  \BibitemOpen
  \bibfield  {author} {\bibinfo {author} {\bibfnamefont {T.}~\bibnamefont
  {Jungwirth}}, \bibinfo {author} {\bibfnamefont {Q.}~\bibnamefont {Niu}},\
  and\ \bibinfo {author} {\bibfnamefont {A.~H.}\ \bibnamefont {MacDonald}},\
  }\bibfield  {title} {\bibinfo {title} {Anomalous {Hall} {Effect} in
  {Ferromagnetic} {Semiconductors}},\ }\href
  {https://doi.org/10.1103/PhysRevLett.88.207208} {\bibfield  {journal}
  {\bibinfo  {journal} {Phys. Rev. Lett.}\ }\textbf {\bibinfo {volume} {88}},\
  \bibinfo {pages} {207208} (\bibinfo {year} {2002})}\BibitemShut {NoStop}%
\bibitem [{\citenamefont {Onoda}\ and\ \citenamefont
  {Nagaosa}(2002)}]{onoda_topological_2002}%
  \BibitemOpen
  \bibfield  {author} {\bibinfo {author} {\bibfnamefont {M.}~\bibnamefont
  {Onoda}}\ and\ \bibinfo {author} {\bibfnamefont {N.}~\bibnamefont
  {Nagaosa}},\ }\bibfield  {title} {\bibinfo {title} {Topological {Nature} of
  {Anomalous} {Hall} {Effect} in {Ferromagnets}},\ }\href
  {https://doi.org/10.1143/JPSJ.71.19} {\bibfield  {journal} {\bibinfo
  {journal} {J. Phys. Soc. Jpn.}\ }\textbf {\bibinfo {volume} {71}},\ \bibinfo
  {pages} {19} (\bibinfo {year} {2002})}\BibitemShut {NoStop}%
\bibitem [{\citenamefont {Berry}(1984)}]{berry1984quantal}%
  \BibitemOpen
  \bibfield  {author} {\bibinfo {author} {\bibfnamefont {M.~V.}\ \bibnamefont
  {Berry}},\ }\bibfield  {title} {\bibinfo {title} {Quantal phase factors
  accompanying adiabatic changes},\ }\href@noop {} {\bibfield  {journal}
  {\bibinfo  {journal} {Proceedings of the Royal Society of London. A.
  Mathematical and Physical Sciences}\ }\textbf {\bibinfo {volume} {392}},\
  \bibinfo {pages} {45} (\bibinfo {year} {1984})}\BibitemShut {NoStop}%
\bibitem [{\citenamefont {Provost}\ and\ \citenamefont
  {Vallee}(1980)}]{provost_riemannian_1980}%
  \BibitemOpen
  \bibfield  {author} {\bibinfo {author} {\bibfnamefont {J.~P.}\ \bibnamefont
  {Provost}}\ and\ \bibinfo {author} {\bibfnamefont {G.}~\bibnamefont
  {Vallee}},\ }\bibfield  {title} {\bibinfo {title} {Riemannian structure on
  manifolds of quantum states},\ }\href {https://doi.org/10.1007/BF02193559}
  {\bibfield  {journal} {\bibinfo  {journal} {Commun.Math. Phys.}\ }\textbf
  {\bibinfo {volume} {76}},\ \bibinfo {pages} {289} (\bibinfo {year}
  {1980})}\BibitemShut {NoStop}%
\bibitem [{\citenamefont {Aversa}\ and\ \citenamefont
  {Sipe}(1995)}]{aversa_nonlinear_1995}%
  \BibitemOpen
  \bibfield  {author} {\bibinfo {author} {\bibfnamefont {C.}~\bibnamefont
  {Aversa}}\ and\ \bibinfo {author} {\bibfnamefont {J.~E.}\ \bibnamefont
  {Sipe}},\ }\bibfield  {title} {\bibinfo {title} {Nonlinear optical
  susceptibilities of semiconductors: {Results} with a length-gauge analysis},\
  }\href {https://doi.org/10.1103/PhysRevB.52.14636} {\bibfield  {journal}
  {\bibinfo  {journal} {Phys. Rev. B}\ }\textbf {\bibinfo {volume} {52}},\
  \bibinfo {pages} {14636} (\bibinfo {year} {1995})}\BibitemShut {NoStop}%
\bibitem [{\citenamefont {Sipe}\ and\ \citenamefont
  {Shkrebtii}(2000)}]{sipe_second-order_2000}%
  \BibitemOpen
  \bibfield  {author} {\bibinfo {author} {\bibfnamefont {J.~E.}\ \bibnamefont
  {Sipe}}\ and\ \bibinfo {author} {\bibfnamefont {A.~I.}\ \bibnamefont
  {Shkrebtii}},\ }\bibfield  {title} {\bibinfo {title} {Second-order optical
  response in semiconductors},\ }\href
  {https://doi.org/10.1103/PhysRevB.61.5337} {\bibfield  {journal} {\bibinfo
  {journal} {Phys. Rev. B}\ }\textbf {\bibinfo {volume} {61}},\ \bibinfo
  {pages} {5337} (\bibinfo {year} {2000})}\BibitemShut {NoStop}%
\bibitem [{\citenamefont {Hosur}(2011)}]{hosur_circular_2011}%
  \BibitemOpen
  \bibfield  {author} {\bibinfo {author} {\bibfnamefont {P.}~\bibnamefont
  {Hosur}},\ }\bibfield  {title} {\bibinfo {title} {Circular photogalvanic
  effect on topological insulator surfaces: {Berry}-curvature-dependent
  response},\ }\href {https://doi.org/10.1103/PhysRevB.83.035309} {\bibfield
  {journal} {\bibinfo  {journal} {Phys. Rev. B}\ }\textbf {\bibinfo {volume}
  {83}},\ \bibinfo {pages} {035309} (\bibinfo {year} {2011})}\BibitemShut
  {NoStop}%
\bibitem [{\citenamefont {Sodemann}\ and\ \citenamefont
  {Fu}(2015)}]{sodemann_quantum_2015}%
  \BibitemOpen
  \bibfield  {author} {\bibinfo {author} {\bibfnamefont {I.}~\bibnamefont
  {Sodemann}}\ and\ \bibinfo {author} {\bibfnamefont {L.}~\bibnamefont {Fu}},\
  }\bibfield  {title} {\bibinfo {title} {Quantum {Nonlinear} {Hall} {Effect}
  {Induced} by {Berry} {Curvature} {Dipole} in {Time}-{Reversal} {Invariant}
  {Materials}},\ }\href {https://doi.org/10.1103/PhysRevLett.115.216806}
  {\bibfield  {journal} {\bibinfo  {journal} {Phys. Rev. Lett.}\ }\textbf
  {\bibinfo {volume} {115}},\ \bibinfo {pages} {216806} (\bibinfo {year}
  {2015})}\BibitemShut {NoStop}%
\bibitem [{\citenamefont {Morimoto}\ \emph {et~al.}(2016)\citenamefont
  {Morimoto}, \citenamefont {Zhong}, \citenamefont {Orenstein},\ and\
  \citenamefont {Moore}}]{morimoto_semiclassical_2016}%
  \BibitemOpen
  \bibfield  {author} {\bibinfo {author} {\bibfnamefont {T.}~\bibnamefont
  {Morimoto}}, \bibinfo {author} {\bibfnamefont {S.}~\bibnamefont {Zhong}},
  \bibinfo {author} {\bibfnamefont {J.}~\bibnamefont {Orenstein}},\ and\
  \bibinfo {author} {\bibfnamefont {J.~E.}\ \bibnamefont {Moore}},\ }\bibfield
  {title} {\bibinfo {title} {Semiclassical theory of nonlinear magneto-optical
  responses with applications to topological {Dirac}/{Weyl} semimetals},\
  }\href {https://doi.org/10.1103/PhysRevB.94.245121} {\bibfield  {journal}
  {\bibinfo  {journal} {Phys. Rev. B}\ }\textbf {\bibinfo {volume} {94}},\
  \bibinfo {pages} {245121} (\bibinfo {year} {2016})}\BibitemShut {NoStop}%
\bibitem [{\citenamefont {Ventura}\ \emph {et~al.}(2017)\citenamefont
  {Ventura}, \citenamefont {Passos}, \citenamefont {Lopes Dos~Santos},
  \citenamefont {Viana Parente~Lopes},\ and\ \citenamefont
  {Peres}}]{ventura_gauge_2017}%
  \BibitemOpen
  \bibfield  {author} {\bibinfo {author} {\bibfnamefont {G.~B.}\ \bibnamefont
  {Ventura}}, \bibinfo {author} {\bibfnamefont {D.~J.}\ \bibnamefont {Passos}},
  \bibinfo {author} {\bibfnamefont {J.~M.~B.}\ \bibnamefont {Lopes
  Dos~Santos}}, \bibinfo {author} {\bibfnamefont {J.~M.}\ \bibnamefont {Viana
  Parente~Lopes}},\ and\ \bibinfo {author} {\bibfnamefont {N.~M.~R.}\
  \bibnamefont {Peres}},\ }\bibfield  {title} {\bibinfo {title} {Gauge
  covariances and nonlinear optical responses},\ }\href
  {https://doi.org/10.1103/PhysRevB.96.035431} {\bibfield  {journal} {\bibinfo
  {journal} {Phys. Rev. B}\ }\textbf {\bibinfo {volume} {96}},\ \bibinfo
  {pages} {035431} (\bibinfo {year} {2017})}\BibitemShut {NoStop}%
\bibitem [{\citenamefont {Parker}\ \emph {et~al.}(2019)\citenamefont {Parker},
  \citenamefont {Morimoto}, \citenamefont {Orenstein},\ and\ \citenamefont
  {Moore}}]{parker_diagrammatic_2019}%
  \BibitemOpen
  \bibfield  {author} {\bibinfo {author} {\bibfnamefont {D.~E.}\ \bibnamefont
  {Parker}}, \bibinfo {author} {\bibfnamefont {T.}~\bibnamefont {Morimoto}},
  \bibinfo {author} {\bibfnamefont {J.}~\bibnamefont {Orenstein}},\ and\
  \bibinfo {author} {\bibfnamefont {J.~E.}\ \bibnamefont {Moore}},\ }\bibfield
  {title} {\bibinfo {title} {Diagrammatic approach to nonlinear optical
  response with application to {Weyl} semimetals},\ }\href
  {https://doi.org/10.1103/PhysRevB.99.045121} {\bibfield  {journal} {\bibinfo
  {journal} {Phys. Rev. B}\ }\textbf {\bibinfo {volume} {99}},\ \bibinfo
  {pages} {045121} (\bibinfo {year} {2019})}\BibitemShut {NoStop}%
\bibitem [{\citenamefont {Ahn}\ \emph {et~al.}(2020)\citenamefont {Ahn},
  \citenamefont {Guo},\ and\ \citenamefont {Nagaosa}}]{ahn_low-frequency_2020}%
  \BibitemOpen
  \bibfield  {author} {\bibinfo {author} {\bibfnamefont {J.}~\bibnamefont
  {Ahn}}, \bibinfo {author} {\bibfnamefont {G.-Y.}\ \bibnamefont {Guo}},\ and\
  \bibinfo {author} {\bibfnamefont {N.}~\bibnamefont {Nagaosa}},\ }\bibfield
  {title} {\bibinfo {title} {Low-{Frequency} {Divergence} and {Quantum}
  {Geometry} of the {Bulk} {Photovoltaic} {Effect} in {Topological}
  {Semimetals}},\ }\href {https://doi.org/10.1103/PhysRevX.10.041041}
  {\bibfield  {journal} {\bibinfo  {journal} {Phys. Rev. X}\ }\textbf {\bibinfo
  {volume} {10}},\ \bibinfo {pages} {041041} (\bibinfo {year}
  {2020})}\BibitemShut {NoStop}%
\bibitem [{\citenamefont {Watanabe}\ and\ \citenamefont
  {Yanase}(2021)}]{watanabe_chiral_2021}%
  \BibitemOpen
  \bibfield  {author} {\bibinfo {author} {\bibfnamefont {H.}~\bibnamefont
  {Watanabe}}\ and\ \bibinfo {author} {\bibfnamefont {Y.}~\bibnamefont
  {Yanase}},\ }\bibfield  {title} {\bibinfo {title} {Chiral {Photocurrent} in
  {Parity}-{Violating} {Magnet} and {Enhanced} {Response} in {Topological}
  {Antiferromagnet}},\ }\href {https://doi.org/10.1103/PhysRevX.11.011001}
  {\bibfield  {journal} {\bibinfo  {journal} {Phys. Rev. X}\ }\textbf {\bibinfo
  {volume} {11}},\ \bibinfo {pages} {011001} (\bibinfo {year}
  {2021})}\BibitemShut {NoStop}%
\bibitem [{\citenamefont {Von~Baltz}\ and\ \citenamefont
  {Kraut}(1981)}]{von_baltz_theory_1981}%
  \BibitemOpen
  \bibfield  {author} {\bibinfo {author} {\bibfnamefont {R.}~\bibnamefont
  {Von~Baltz}}\ and\ \bibinfo {author} {\bibfnamefont {W.}~\bibnamefont
  {Kraut}},\ }\bibfield  {title} {\bibinfo {title} {Theory of the bulk
  photovoltaic effect in pure crystals},\ }\href
  {https://doi.org/10.1103/PhysRevB.23.5590} {\bibfield  {journal} {\bibinfo
  {journal} {Phys. Rev. B}\ }\textbf {\bibinfo {volume} {23}},\ \bibinfo
  {pages} {5590} (\bibinfo {year} {1981})}\BibitemShut {NoStop}%
\bibitem [{\citenamefont {Fridkin}(2001)}]{fridkin_bulk_2001}%
  \BibitemOpen
  \bibfield  {author} {\bibinfo {author} {\bibfnamefont {V.~M.}\ \bibnamefont
  {Fridkin}},\ }\bibfield  {title} {\bibinfo {title} {Bulk photovoltaic effect
  in noncentrosymmetric crystals},\ }\href {https://doi.org/10.1134/1.1387133}
  {\bibfield  {journal} {\bibinfo  {journal} {Crystallogr. Rep.}\ }\textbf
  {\bibinfo {volume} {46}},\ \bibinfo {pages} {654} (\bibinfo {year}
  {2001})}\BibitemShut {NoStop}%
\bibitem [{\citenamefont {Sturman}\ and\ \citenamefont
  {Fridkin}(2021)}]{sturman_photovoltaic_2021}%
  \BibitemOpen
  \bibfield  {author} {\bibinfo {author} {\bibfnamefont {B.~I.}\ \bibnamefont
  {Sturman}}\ and\ \bibinfo {author} {\bibfnamefont {V.~M.}\ \bibnamefont
  {Fridkin}},\ }\href {https://doi.org/10.1201/9780203743416} {\emph {\bibinfo
  {title} {The {Photovoltaic} and {Photorefractive} {Effects} in
  {Noncentrosymmetric} {Materials}}}},\ \bibinfo {edition} {1st}\ ed.\
  (\bibinfo  {publisher} {Routledge, London},\ \bibinfo {year}
  {2021})\BibitemShut {NoStop}%
\bibitem [{\citenamefont {Ribakovs}\ and\ \citenamefont
  {Gundjian}(1977)}]{ribakovs_theory_1977}%
  \BibitemOpen
  \bibfield  {author} {\bibinfo {author} {\bibfnamefont {G.}~\bibnamefont
  {Ribakovs}}\ and\ \bibinfo {author} {\bibfnamefont {A.~A.}\ \bibnamefont
  {Gundjian}},\ }\bibfield  {title} {\bibinfo {title} {Theory of the photon
  drag effect in tellurium},\ }\href {https://doi.org/10.1063/1.323520}
  {\bibfield  {journal} {\bibinfo  {journal} {Journal of Applied Physics}\
  }\textbf {\bibinfo {volume} {48}},\ \bibinfo {pages} {4609} (\bibinfo {year}
  {1977})}\BibitemShut {NoStop}%
\bibitem [{\citenamefont {Goff}\ and\ \citenamefont
  {Schaich}(2000)}]{goff_theory_2000}%
  \BibitemOpen
  \bibfield  {author} {\bibinfo {author} {\bibfnamefont {J.~E.}\ \bibnamefont
  {Goff}}\ and\ \bibinfo {author} {\bibfnamefont {W.~L.}\ \bibnamefont
  {Schaich}},\ }\bibfield  {title} {\bibinfo {title} {Theory of the photon-drag
  effect in simple metals},\ }\href {https://doi.org/10.1103/PhysRevB.61.10471}
  {\bibfield  {journal} {\bibinfo  {journal} {Phys. Rev. B}\ }\textbf {\bibinfo
  {volume} {61}},\ \bibinfo {pages} {10471} (\bibinfo {year}
  {2000})}\BibitemShut {NoStop}%
\bibitem [{\citenamefont {Entin}\ \emph {et~al.}(2010)\citenamefont {Entin},
  \citenamefont {Magarill},\ and\ \citenamefont
  {Shepelyansky}}]{entin_theory_2010}%
  \BibitemOpen
  \bibfield  {author} {\bibinfo {author} {\bibfnamefont {M.~V.}\ \bibnamefont
  {Entin}}, \bibinfo {author} {\bibfnamefont {L.~I.}\ \bibnamefont
  {Magarill}},\ and\ \bibinfo {author} {\bibfnamefont {D.~L.}\ \bibnamefont
  {Shepelyansky}},\ }\bibfield  {title} {\bibinfo {title} {Theory of resonant
  photon drag in monolayer graphene},\ }\href
  {https://doi.org/10.1103/PhysRevB.81.165441} {\bibfield  {journal} {\bibinfo
  {journal} {Phys. Rev. B}\ }\textbf {\bibinfo {volume} {81}},\ \bibinfo
  {pages} {165441} (\bibinfo {year} {2010})}\BibitemShut {NoStop}%
\bibitem [{\citenamefont {Shi}\ \emph {et~al.}(2021)\citenamefont {Shi},
  \citenamefont {Zhang}, \citenamefont {Chang},\ and\ \citenamefont
  {Song}}]{shi_geometric_2021}%
  \BibitemOpen
  \bibfield  {author} {\bibinfo {author} {\bibfnamefont {L.-k.}\ \bibnamefont
  {Shi}}, \bibinfo {author} {\bibfnamefont {D.}~\bibnamefont {Zhang}}, \bibinfo
  {author} {\bibfnamefont {K.}~\bibnamefont {Chang}},\ and\ \bibinfo {author}
  {\bibfnamefont {J.~C.}\ \bibnamefont {Song}},\ }\bibfield  {title} {\bibinfo
  {title} {Geometric {Photon}-{Drag} {Effect} and {Nonlinear} {Shift} {Current}
  in {Centrosymmetric} {Crystals}},\ }\href
  {https://doi.org/10.1103/PhysRevLett.126.197402} {\bibfield  {journal}
  {\bibinfo  {journal} {Phys. Rev. Lett.}\ }\textbf {\bibinfo {volume} {126}},\
  \bibinfo {pages} {197402} (\bibinfo {year} {2021})}\BibitemShut {NoStop}%
\bibitem [{\citenamefont {Basov}\ \emph {et~al.}(2016)\citenamefont {Basov},
  \citenamefont {Fogler},\ and\ \citenamefont {García
  De~Abajo}}]{basov_polaritons_2016}%
  \BibitemOpen
  \bibfield  {author} {\bibinfo {author} {\bibfnamefont {D.~N.}\ \bibnamefont
  {Basov}}, \bibinfo {author} {\bibfnamefont {M.~M.}\ \bibnamefont {Fogler}},\
  and\ \bibinfo {author} {\bibfnamefont {F.~J.}\ \bibnamefont {García
  De~Abajo}},\ }\bibfield  {title} {\bibinfo {title} {Polaritons in van der
  {Waals} materials},\ }\href {https://doi.org/10.1126/science.aag1992}
  {\bibfield  {journal} {\bibinfo  {journal} {Science}\ }\textbf {\bibinfo
  {volume} {354}},\ \bibinfo {pages} {aag1992} (\bibinfo {year}
  {2016})}\BibitemShut {NoStop}%
\bibitem [{\citenamefont {Kurman}\ \emph {et~al.}(2018)\citenamefont {Kurman},
  \citenamefont {Rivera}, \citenamefont {Christensen}, \citenamefont {Tsesses},
  \citenamefont {Orenstein}, \citenamefont {Soljačić}, \citenamefont
  {Joannopoulos},\ and\ \citenamefont {Kaminer}}]{kurman_control_2018}%
  \BibitemOpen
  \bibfield  {author} {\bibinfo {author} {\bibfnamefont {Y.}~\bibnamefont
  {Kurman}}, \bibinfo {author} {\bibfnamefont {N.}~\bibnamefont {Rivera}},
  \bibinfo {author} {\bibfnamefont {T.}~\bibnamefont {Christensen}}, \bibinfo
  {author} {\bibfnamefont {S.}~\bibnamefont {Tsesses}}, \bibinfo {author}
  {\bibfnamefont {M.}~\bibnamefont {Orenstein}}, \bibinfo {author}
  {\bibfnamefont {M.}~\bibnamefont {Soljačić}}, \bibinfo {author}
  {\bibfnamefont {J.~D.}\ \bibnamefont {Joannopoulos}},\ and\ \bibinfo {author}
  {\bibfnamefont {I.}~\bibnamefont {Kaminer}},\ }\bibfield  {title} {\bibinfo
  {title} {Control of semiconductor emitter frequency by increasing polariton
  momenta},\ }\href {https://doi.org/10.1038/s41566-018-0176-6} {\bibfield
  {journal} {\bibinfo  {journal} {Nature Photon}\ }\textbf {\bibinfo {volume}
  {12}},\ \bibinfo {pages} {423} (\bibinfo {year} {2018})}\BibitemShut
  {NoStop}%
\bibitem [{\citenamefont {Hatano}\ \emph {et~al.}(2009)\citenamefont {Hatano},
  \citenamefont {Ishihara}, \citenamefont {Tikhodeev},\ and\ \citenamefont
  {Gippius}}]{hatano_transverse_2009}%
  \BibitemOpen
  \bibfield  {author} {\bibinfo {author} {\bibfnamefont {T.}~\bibnamefont
  {Hatano}}, \bibinfo {author} {\bibfnamefont {T.}~\bibnamefont {Ishihara}},
  \bibinfo {author} {\bibfnamefont {S.~G.}\ \bibnamefont {Tikhodeev}},\ and\
  \bibinfo {author} {\bibfnamefont {N.~A.}\ \bibnamefont {Gippius}},\
  }\bibfield  {title} {\bibinfo {title} {Transverse {Photovoltage} {Induced} by
  {Circularly} {Polarized} {Light}},\ }\href
  {https://doi.org/10.1103/PhysRevLett.103.103906} {\bibfield  {journal}
  {\bibinfo  {journal} {Phys. Rev. Lett.}\ }\textbf {\bibinfo {volume} {103}},\
  \bibinfo {pages} {103906} (\bibinfo {year} {2009})}\BibitemShut {NoStop}%
\bibitem [{\citenamefont {Karch}\ \emph {et~al.}(2010)\citenamefont {Karch},
  \citenamefont {Olbrich}, \citenamefont {Schmalzbauer}, \citenamefont {Zoth},
  \citenamefont {Brinsteiner}, \citenamefont {Fehrenbacher}, \citenamefont
  {Wurstbauer}, \citenamefont {Glazov}, \citenamefont {Tarasenko},
  \citenamefont {Ivchenko}, \citenamefont {Weiss}, \citenamefont {Eroms},
  \citenamefont {Yakimova}, \citenamefont {Lara-Avila}, \citenamefont
  {Kubatkin},\ and\ \citenamefont {Ganichev}}]{karch_dynamic_2010}%
  \BibitemOpen
  \bibfield  {author} {\bibinfo {author} {\bibfnamefont {J.}~\bibnamefont
  {Karch}}, \bibinfo {author} {\bibfnamefont {P.}~\bibnamefont {Olbrich}},
  \bibinfo {author} {\bibfnamefont {M.}~\bibnamefont {Schmalzbauer}}, \bibinfo
  {author} {\bibfnamefont {C.}~\bibnamefont {Zoth}}, \bibinfo {author}
  {\bibfnamefont {C.}~\bibnamefont {Brinsteiner}}, \bibinfo {author}
  {\bibfnamefont {M.}~\bibnamefont {Fehrenbacher}}, \bibinfo {author}
  {\bibfnamefont {U.}~\bibnamefont {Wurstbauer}}, \bibinfo {author}
  {\bibfnamefont {M.~M.}\ \bibnamefont {Glazov}}, \bibinfo {author}
  {\bibfnamefont {S.~A.}\ \bibnamefont {Tarasenko}}, \bibinfo {author}
  {\bibfnamefont {E.~L.}\ \bibnamefont {Ivchenko}}, \bibinfo {author}
  {\bibfnamefont {D.}~\bibnamefont {Weiss}}, \bibinfo {author} {\bibfnamefont
  {J.}~\bibnamefont {Eroms}}, \bibinfo {author} {\bibfnamefont
  {R.}~\bibnamefont {Yakimova}}, \bibinfo {author} {\bibfnamefont
  {S.}~\bibnamefont {Lara-Avila}}, \bibinfo {author} {\bibfnamefont
  {S.}~\bibnamefont {Kubatkin}},\ and\ \bibinfo {author} {\bibfnamefont
  {S.~D.}\ \bibnamefont {Ganichev}},\ }\bibfield  {title} {\bibinfo {title}
  {Dynamic {Hall} {Effect} {Driven} by {Circularly} {Polarized} {Light} in a
  {Graphene} {Layer}},\ }\href {https://doi.org/10.1103/PhysRevLett.105.227402}
  {\bibfield  {journal} {\bibinfo  {journal} {Phys. Rev. Lett.}\ }\textbf
  {\bibinfo {volume} {105}},\ \bibinfo {pages} {227402} (\bibinfo {year}
  {2010})}\BibitemShut {NoStop}%
\bibitem [{\citenamefont {Shalygin}\ \emph {et~al.}(2016)\citenamefont
  {Shalygin}, \citenamefont {Moldavskaya}, \citenamefont {Danilov},
  \citenamefont {Farbshtein},\ and\ \citenamefont
  {Golub}}]{shalygin_circular_2016}%
  \BibitemOpen
  \bibfield  {author} {\bibinfo {author} {\bibfnamefont {V.~A.}\ \bibnamefont
  {Shalygin}}, \bibinfo {author} {\bibfnamefont {M.~D.}\ \bibnamefont
  {Moldavskaya}}, \bibinfo {author} {\bibfnamefont {S.~N.}\ \bibnamefont
  {Danilov}}, \bibinfo {author} {\bibfnamefont {I.~I.}\ \bibnamefont
  {Farbshtein}},\ and\ \bibinfo {author} {\bibfnamefont {L.~E.}\ \bibnamefont
  {Golub}},\ }\bibfield  {title} {\bibinfo {title} {Circular photon drag effect
  in bulk tellurium},\ }\href {https://doi.org/10.1103/PhysRevB.93.045207}
  {\bibfield  {journal} {\bibinfo  {journal} {Phys. Rev. B}\ }\textbf {\bibinfo
  {volume} {93}},\ \bibinfo {pages} {045207} (\bibinfo {year}
  {2016})}\BibitemShut {NoStop}%
\bibitem [{\citenamefont {Strait}\ \emph {et~al.}(2019)\citenamefont {Strait},
  \citenamefont {Holland}, \citenamefont {Zhu}, \citenamefont {Zhang},
  \citenamefont {Ilic}, \citenamefont {Agrawal}, \citenamefont {Pacifici},\
  and\ \citenamefont {Lezec}}]{strait_revisiting_2019}%
  \BibitemOpen
  \bibfield  {author} {\bibinfo {author} {\bibfnamefont {J.~H.}\ \bibnamefont
  {Strait}}, \bibinfo {author} {\bibfnamefont {G.}~\bibnamefont {Holland}},
  \bibinfo {author} {\bibfnamefont {W.}~\bibnamefont {Zhu}}, \bibinfo {author}
  {\bibfnamefont {C.}~\bibnamefont {Zhang}}, \bibinfo {author} {\bibfnamefont
  {B.~R.}\ \bibnamefont {Ilic}}, \bibinfo {author} {\bibfnamefont
  {A.}~\bibnamefont {Agrawal}}, \bibinfo {author} {\bibfnamefont
  {D.}~\bibnamefont {Pacifici}},\ and\ \bibinfo {author} {\bibfnamefont
  {H.~J.}\ \bibnamefont {Lezec}},\ }\bibfield  {title} {\bibinfo {title}
  {Revisiting the {Photon}-{Drag} {Effect} in {Metal} {Films}},\ }\href
  {https://doi.org/10.1103/PhysRevLett.123.053903} {\bibfield  {journal}
  {\bibinfo  {journal} {Phys. Rev. Lett.}\ }\textbf {\bibinfo {volume} {123}},\
  \bibinfo {pages} {053903} (\bibinfo {year} {2019})}\BibitemShut {NoStop}%
\bibitem [{\citenamefont {Shalygin}\ \emph {et~al.}(2007)\citenamefont
  {Shalygin}, \citenamefont {Diehl}, \citenamefont {Hoffmann}, \citenamefont
  {Danilov}, \citenamefont {Herrle}, \citenamefont {Tarasenko}, \citenamefont
  {Schuh}, \citenamefont {Gerl}, \citenamefont {Wegscheider}, \citenamefont
  {Prettl},\ and\ \citenamefont {Ganichev}}]{shalygin_spin_2007}%
  \BibitemOpen
  \bibfield  {author} {\bibinfo {author} {\bibfnamefont {V.~A.}\ \bibnamefont
  {Shalygin}}, \bibinfo {author} {\bibfnamefont {H.}~\bibnamefont {Diehl}},
  \bibinfo {author} {\bibfnamefont {C.}~\bibnamefont {Hoffmann}}, \bibinfo
  {author} {\bibfnamefont {S.~N.}\ \bibnamefont {Danilov}}, \bibinfo {author}
  {\bibfnamefont {T.}~\bibnamefont {Herrle}}, \bibinfo {author} {\bibfnamefont
  {S.~A.}\ \bibnamefont {Tarasenko}}, \bibinfo {author} {\bibfnamefont
  {D.}~\bibnamefont {Schuh}}, \bibinfo {author} {\bibfnamefont
  {C.}~\bibnamefont {Gerl}}, \bibinfo {author} {\bibfnamefont {W.}~\bibnamefont
  {Wegscheider}}, \bibinfo {author} {\bibfnamefont {W.}~\bibnamefont
  {Prettl}},\ and\ \bibinfo {author} {\bibfnamefont {S.~D.}\ \bibnamefont
  {Ganichev}},\ }\bibfield  {title} {\bibinfo {title} {Spin photocurrents and
  the circular photon drag effect in (110)-grown quantum well structures},\
  }\href {https://doi.org/10.1134/S0021364006220097} {\bibfield  {journal}
  {\bibinfo  {journal} {Jetp Lett.}\ }\textbf {\bibinfo {volume} {84}},\
  \bibinfo {pages} {570} (\bibinfo {year} {2007})}\BibitemShut {NoStop}%
\bibitem [{\citenamefont {Hamara}\ \emph {et~al.}(2023)\citenamefont {Hamara},
  \citenamefont {Lange}, \citenamefont {Kholid}, \citenamefont {Markou},
  \citenamefont {Felser}, \citenamefont {Slager},\ and\ \citenamefont
  {Ciccarelli}}]{hamara_ultrafast_2023}%
  \BibitemOpen
  \bibfield  {author} {\bibinfo {author} {\bibfnamefont {D.}~\bibnamefont
  {Hamara}}, \bibinfo {author} {\bibfnamefont {G.~F.}\ \bibnamefont {Lange}},
  \bibinfo {author} {\bibfnamefont {F.~N.}\ \bibnamefont {Kholid}}, \bibinfo
  {author} {\bibfnamefont {A.}~\bibnamefont {Markou}}, \bibinfo {author}
  {\bibfnamefont {C.}~\bibnamefont {Felser}}, \bibinfo {author} {\bibfnamefont
  {R.-J.}\ \bibnamefont {Slager}},\ and\ \bibinfo {author} {\bibfnamefont
  {C.}~\bibnamefont {Ciccarelli}},\ }\bibfield  {title} {\bibinfo {title}
  {Ultrafast helicity-dependent photocurrents in {Weyl} {Magnet} {Mn3Sn}},\
  }\href {https://doi.org/10.1038/s42005-023-01440-5} {\bibfield  {journal}
  {\bibinfo  {journal} {Commun. Phys.}\ }\textbf {\bibinfo {volume} {6}},\
  \bibinfo {pages} {320} (\bibinfo {year} {2023})}\BibitemShut {NoStop}%
\bibitem [{\citenamefont {Fuseya}\ \emph {et~al.}(2015)\citenamefont {Fuseya},
  \citenamefont {Ogata},\ and\ \citenamefont {Fukuyama}}]{fuseya2015transport}%
  \BibitemOpen
  \bibfield  {author} {\bibinfo {author} {\bibfnamefont {Y.}~\bibnamefont
  {Fuseya}}, \bibinfo {author} {\bibfnamefont {M.}~\bibnamefont {Ogata}},\ and\
  \bibinfo {author} {\bibfnamefont {H.}~\bibnamefont {Fukuyama}},\ }\bibfield
  {title} {\bibinfo {title} {Transport properties and diamagnetism of dirac
  electrons in bismuth},\ }\href {https://doi.org/10.7566/JPSJ.84.012001}
  {\bibfield  {journal} {\bibinfo  {journal} {J. Phys. Soc. Jpn.}\ }\textbf
  {\bibinfo {volume} {84}},\ \bibinfo {pages} {012001} (\bibinfo {year}
  {2015})}\BibitemShut {NoStop}%
\bibitem [{\citenamefont {Chi}\ \emph {et~al.}(2022)\citenamefont {Chi},
  \citenamefont {Qu}, \citenamefont {Lau}, \citenamefont {Kawaguchi},
  \citenamefont {Fujimoto}, \citenamefont {Takanashi}, \citenamefont {Ogata},\
  and\ \citenamefont {Hayashi}}]{chi_spin_2022}%
  \BibitemOpen
  \bibfield  {author} {\bibinfo {author} {\bibfnamefont {Z.}~\bibnamefont
  {Chi}}, \bibinfo {author} {\bibfnamefont {G.}~\bibnamefont {Qu}}, \bibinfo
  {author} {\bibfnamefont {Y.-C.}\ \bibnamefont {Lau}}, \bibinfo {author}
  {\bibfnamefont {M.}~\bibnamefont {Kawaguchi}}, \bibinfo {author}
  {\bibfnamefont {J.}~\bibnamefont {Fujimoto}}, \bibinfo {author}
  {\bibfnamefont {K.}~\bibnamefont {Takanashi}}, \bibinfo {author}
  {\bibfnamefont {M.}~\bibnamefont {Ogata}},\ and\ \bibinfo {author}
  {\bibfnamefont {M.}~\bibnamefont {Hayashi}},\ }\bibfield  {title} {\bibinfo
  {title} {Spin {Hall} effect driven by the spin magnetic moment current in
  {Dirac} materials},\ }\href {https://doi.org/10.1103/PhysRevB.105.214419}
  {\bibfield  {journal} {\bibinfo  {journal} {Phys. Rev. B}\ }\textbf {\bibinfo
  {volume} {105}},\ \bibinfo {pages} {214419} (\bibinfo {year}
  {2022})}\BibitemShut {NoStop}%
\bibitem [{sup()}]{supplementary}%
  \BibitemOpen
  \href@noop {} {}\bibinfo {note} {See Supplemental Material at
  URL-will-be-inserted-by-publisher for the data of the
  experiments.}\BibitemShut {Stop}%
\bibitem [{Note1()}]{Note1}%
  \BibitemOpen
  \bibinfo {note} {The explicit form of covariant derivative on degenerate
  bands is $[\protect \mathcal {D}_\gamma O ]_{nm} = \partial _{\gamma } O_{nm}
  - i \left ( \DOTSB \sum@ \slimits@ \limits _{\protect \tilde {l} = \protect
  \tilde {n}} A^\gamma _{nl} O_{lm} - \DOTSB \sum@ \slimits@ \limits _{\protect
  \tilde {l} = \protect \tilde {m}} O_{nl} A^\gamma _{lm} \right )$ with
  summation over degenerate subspaces $\protect \tilde {l}=\protect \tilde
  {n},\protect \tilde {m}$.}\BibitemShut {Stop}%
\bibitem [{\citenamefont {Fregoso}(2019)}]{PhysRevB.100.064301}%
  \BibitemOpen
  \bibfield  {author} {\bibinfo {author} {\bibfnamefont {B.~M.}\ \bibnamefont
  {Fregoso}},\ }\bibfield  {title} {\bibinfo {title} {Bulk photovoltaic effects
  in the presence of a static electric field},\ }\href
  {https://doi.org/10.1103/PhysRevB.100.064301} {\bibfield  {journal} {\bibinfo
   {journal} {Phys. Rev. B}\ }\textbf {\bibinfo {volume} {100}},\ \bibinfo
  {pages} {064301} (\bibinfo {year} {2019})}\BibitemShut {NoStop}%
\bibitem [{\citenamefont {Du}\ \emph {et~al.}(2019)\citenamefont {Du},
  \citenamefont {Wang}, \citenamefont {Li}, \citenamefont {Lu},\ and\
  \citenamefont {Xie}}]{du2019disorder}%
  \BibitemOpen
  \bibfield  {author} {\bibinfo {author} {\bibfnamefont {Z.}~\bibnamefont
  {Du}}, \bibinfo {author} {\bibfnamefont {C.}~\bibnamefont {Wang}}, \bibinfo
  {author} {\bibfnamefont {S.}~\bibnamefont {Li}}, \bibinfo {author}
  {\bibfnamefont {H.-Z.}\ \bibnamefont {Lu}},\ and\ \bibinfo {author}
  {\bibfnamefont {X.}~\bibnamefont {Xie}},\ }\bibfield  {title} {\bibinfo
  {title} {Disorder-induced nonlinear hall effect with time-reversal
  symmetry},\ }\href {https://doi.org/10.1038/s41467-019-10941-3} {\bibfield
  {journal} {\bibinfo  {journal} {Nat. Commun.}\ }\textbf {\bibinfo {volume}
  {10}},\ \bibinfo {pages} {3047} (\bibinfo {year} {2019})}\BibitemShut
  {NoStop}%
\bibitem [{\citenamefont {Blount}(1962)}]{blount1962formalisms}%
  \BibitemOpen
  \bibfield  {author} {\bibinfo {author} {\bibfnamefont {E.}~\bibnamefont
  {Blount}},\ }\bibfield  {title} {\bibinfo {title} {Formalisms of band
  theory}\ }(\bibinfo  {publisher} {Academic Press, New York},\ \bibinfo {year}
  {1962})\ pp.\ \bibinfo {pages} {305--373}\BibitemShut {NoStop}%
\bibitem [{\citenamefont {Xu}\ \emph {et~al.}(2018)\citenamefont {Xu},
  \citenamefont {Ma}, \citenamefont {Shen}, \citenamefont {Fatemi},
  \citenamefont {Wu}, \citenamefont {Chang}, \citenamefont {Chang},
  \citenamefont {Valdivia}, \citenamefont {Chan}, \citenamefont {Gibson} \emph
  {et~al.}}]{xu_electrically_2018}%
  \BibitemOpen
  \bibfield  {author} {\bibinfo {author} {\bibfnamefont {S.-Y.}\ \bibnamefont
  {Xu}}, \bibinfo {author} {\bibfnamefont {Q.}~\bibnamefont {Ma}}, \bibinfo
  {author} {\bibfnamefont {H.}~\bibnamefont {Shen}}, \bibinfo {author}
  {\bibfnamefont {V.}~\bibnamefont {Fatemi}}, \bibinfo {author} {\bibfnamefont
  {S.}~\bibnamefont {Wu}}, \bibinfo {author} {\bibfnamefont {T.-R.}\
  \bibnamefont {Chang}}, \bibinfo {author} {\bibfnamefont {G.}~\bibnamefont
  {Chang}}, \bibinfo {author} {\bibfnamefont {A.~M.~M.}\ \bibnamefont
  {Valdivia}}, \bibinfo {author} {\bibfnamefont {C.-K.}\ \bibnamefont {Chan}},
  \bibinfo {author} {\bibfnamefont {Q.~D.}\ \bibnamefont {Gibson}}, \emph
  {et~al.},\ }\bibfield  {title} {\bibinfo {title} {Electrically switchable
  berry curvature dipole in the monolayer topological insulator wte2},\ }\href
  {https://doi.org/10.1038/s41567-018-0189-6} {\bibfield  {journal} {\bibinfo
  {journal} {Nat. Phys.}\ }\textbf {\bibinfo {volume} {14}},\ \bibinfo {pages}
  {900} (\bibinfo {year} {2018})}\BibitemShut {NoStop}%
\bibitem [{\citenamefont {Duan}\ \emph {et~al.}(2023)\citenamefont {Duan},
  \citenamefont {Qin}, \citenamefont {Chen}, \citenamefont {Yang},
  \citenamefont {Qiu}, \citenamefont {Huang}, \citenamefont {Liu},
  \citenamefont {Li}, \citenamefont {Bi}, \citenamefont {Meng}, \citenamefont
  {Xi}, \citenamefont {Yao}, \citenamefont {Ideue}, \citenamefont {Lian},
  \citenamefont {Iwasa},\ and\ \citenamefont {Yuan}}]{duan_berry_2023}%
  \BibitemOpen
  \bibfield  {author} {\bibinfo {author} {\bibfnamefont {S.}~\bibnamefont
  {Duan}}, \bibinfo {author} {\bibfnamefont {F.}~\bibnamefont {Qin}}, \bibinfo
  {author} {\bibfnamefont {P.}~\bibnamefont {Chen}}, \bibinfo {author}
  {\bibfnamefont {X.}~\bibnamefont {Yang}}, \bibinfo {author} {\bibfnamefont
  {C.}~\bibnamefont {Qiu}}, \bibinfo {author} {\bibfnamefont {J.}~\bibnamefont
  {Huang}}, \bibinfo {author} {\bibfnamefont {G.}~\bibnamefont {Liu}}, \bibinfo
  {author} {\bibfnamefont {Z.}~\bibnamefont {Li}}, \bibinfo {author}
  {\bibfnamefont {X.}~\bibnamefont {Bi}}, \bibinfo {author} {\bibfnamefont
  {F.}~\bibnamefont {Meng}}, \bibinfo {author} {\bibfnamefont {X.}~\bibnamefont
  {Xi}}, \bibinfo {author} {\bibfnamefont {J.}~\bibnamefont {Yao}}, \bibinfo
  {author} {\bibfnamefont {T.}~\bibnamefont {Ideue}}, \bibinfo {author}
  {\bibfnamefont {B.}~\bibnamefont {Lian}}, \bibinfo {author} {\bibfnamefont
  {Y.}~\bibnamefont {Iwasa}},\ and\ \bibinfo {author} {\bibfnamefont
  {H.}~\bibnamefont {Yuan}},\ }\bibfield  {title} {\bibinfo {title} {Berry
  curvature dipole generation and helicity-to-spin conversion at
  symmetry-mismatched heterointerfaces},\ }\href
  {https://doi.org/10.1038/s41565-023-01417-z} {\bibfield  {journal} {\bibinfo
  {journal} {Nat. Nanotechnol.}\ }\textbf {\bibinfo {volume} {18}},\ \bibinfo
  {pages} {867} (\bibinfo {year} {2023})}\BibitemShut {NoStop}%
\bibitem [{\citenamefont {Nakahara}(2003)}]{nakahara_geometry_2003}%
  \BibitemOpen
  \bibfield  {author} {\bibinfo {author} {\bibfnamefont {M.}~\bibnamefont
  {Nakahara}},\ }\href {https://doi.org/10.1201/9781420056945} {\emph {\bibinfo
  {title} {Geometry, {Topology} and {Physics}, {Second} {Edition}}}}\ (\bibinfo
   {publisher} {Taylor \&amp; Francis, Boca Raton},\ \bibinfo {year}
  {2003})\BibitemShut {NoStop}%
\bibitem [{\citenamefont {Gao}\ and\ \citenamefont
  {Xiao}(2019)}]{gao_nonreciprocal_2019}%
  \BibitemOpen
  \bibfield  {author} {\bibinfo {author} {\bibfnamefont {Y.}~\bibnamefont
  {Gao}}\ and\ \bibinfo {author} {\bibfnamefont {D.}~\bibnamefont {Xiao}},\
  }\bibfield  {title} {\bibinfo {title} {Nonreciprocal {Directional}
  {Dichroism} {Induced} by the {Quantum} {Metric} {Dipole}},\ }\href
  {https://doi.org/10.1103/PhysRevLett.122.227402} {\bibfield  {journal}
  {\bibinfo  {journal} {Phys. Rev. Lett.}\ }\textbf {\bibinfo {volume} {122}},\
  \bibinfo {pages} {227402} (\bibinfo {year} {2019})}\BibitemShut {NoStop}%
\bibitem [{\citenamefont {Das}\ \emph {et~al.}(2023)\citenamefont {Das},
  \citenamefont {Lahiri}, \citenamefont {Atencia}, \citenamefont {Culcer},\
  and\ \citenamefont {Agarwal}}]{das_intrinsic_2023}%
  \BibitemOpen
  \bibfield  {author} {\bibinfo {author} {\bibfnamefont {K.}~\bibnamefont
  {Das}}, \bibinfo {author} {\bibfnamefont {S.}~\bibnamefont {Lahiri}},
  \bibinfo {author} {\bibfnamefont {R.~B.}\ \bibnamefont {Atencia}}, \bibinfo
  {author} {\bibfnamefont {D.}~\bibnamefont {Culcer}},\ and\ \bibinfo {author}
  {\bibfnamefont {A.}~\bibnamefont {Agarwal}},\ }\bibfield  {title} {\bibinfo
  {title} {Intrinsic nonlinear conductivities induced by the quantum metric},\
  }\href {https://doi.org/10.1103/PhysRevB.108.L201405} {\bibfield  {journal}
  {\bibinfo  {journal} {Phys. Rev. B}\ }\textbf {\bibinfo {volume} {108}},\
  \bibinfo {pages} {L201405} (\bibinfo {year} {2023})}\BibitemShut {NoStop}%
\bibitem [{\citenamefont {Kaplan}\ \emph {et~al.}(2024)\citenamefont {Kaplan},
  \citenamefont {Holder},\ and\ \citenamefont {Yan}}]{kaplan_unification_2024}%
  \BibitemOpen
  \bibfield  {author} {\bibinfo {author} {\bibfnamefont {D.}~\bibnamefont
  {Kaplan}}, \bibinfo {author} {\bibfnamefont {T.}~\bibnamefont {Holder}},\
  and\ \bibinfo {author} {\bibfnamefont {B.}~\bibnamefont {Yan}},\ }\bibfield
  {title} {\bibinfo {title} {Unification of {Nonlinear} {Anomalous} {Hall}
  {Effect} and {Nonreciprocal} {Magnetoresistance} in {Metals} by the {Quantum}
  {Geometry}},\ }\href {https://doi.org/10.1103/PhysRevLett.132.026301}
  {\bibfield  {journal} {\bibinfo  {journal} {Phys. Rev. Lett.}\ }\textbf
  {\bibinfo {volume} {132}},\ \bibinfo {pages} {026301} (\bibinfo {year}
  {2024})}\BibitemShut {NoStop}%
\bibitem [{\citenamefont {Vecchi}\ and\ \citenamefont
  {Dresselhaus}(1974)}]{PhysRevB.10.771}%
  \BibitemOpen
  \bibfield  {author} {\bibinfo {author} {\bibfnamefont {M.~P.}\ \bibnamefont
  {Vecchi}}\ and\ \bibinfo {author} {\bibfnamefont {M.~S.}\ \bibnamefont
  {Dresselhaus}},\ }\bibfield  {title} {\bibinfo {title} {Temperature
  dependence of the band parameters of bismuth},\ }\href
  {https://doi.org/10.1103/PhysRevB.10.771} {\bibfield  {journal} {\bibinfo
  {journal} {Phys. Rev. B}\ }\textbf {\bibinfo {volume} {10}},\ \bibinfo
  {pages} {771} (\bibinfo {year} {1974})}\BibitemShut {NoStop}%
\bibitem [{\citenamefont {Zhu}\ \emph {et~al.}(2011)\citenamefont {Zhu},
  \citenamefont {Fauqu\'e}, \citenamefont {Fuseya},\ and\ \citenamefont
  {Behnia}}]{PhysRevB.84.115137}%
  \BibitemOpen
  \bibfield  {author} {\bibinfo {author} {\bibfnamefont {Z.}~\bibnamefont
  {Zhu}}, \bibinfo {author} {\bibfnamefont {B.}~\bibnamefont {Fauqu\'e}},
  \bibinfo {author} {\bibfnamefont {Y.}~\bibnamefont {Fuseya}},\ and\ \bibinfo
  {author} {\bibfnamefont {K.}~\bibnamefont {Behnia}},\ }\bibfield  {title}
  {\bibinfo {title} {Angle-resolved landau spectrum of electrons and holes in
  bismuth},\ }\href {https://doi.org/10.1103/PhysRevB.84.115137} {\bibfield
  {journal} {\bibinfo  {journal} {Phys. Rev. B}\ }\textbf {\bibinfo {volume}
  {84}},\ \bibinfo {pages} {115137} (\bibinfo {year} {2011})}\BibitemShut
  {NoStop}%
\end{thebibliography}%

\clearpage
\setcounter{equation}{0}
\renewcommand{\theequation}{S-\arabic{equation}}
\graphicspath {{./Figures/}}
\title{Supplementary material for \textit{Circular Photon Drag Effect in Dirac electrons by Quantum Geometry}}

\author{Guanxiong Qu}
\affiliation{RIKEN Center for Emergent Matter Science (CEMS), Wako 351-0198, Japan.}

\begin{widetext}
\appendix

\begin{center}
    {\large \textbf{Supplemental Material: \\``Density Matrix Renormalization Group Study of Domain Wall Qubits"}}\\[10pt]
    { Guanxiong Qu}\\
    { \textit{ RIKEN, Center for Emergent Matter Science (CEMS), Wako-shi, Saitama 351-0198, Japan}}\\
    {\today}
\end{center}

\section{General Formula for Photon Drag Effect}\label{Sup:A}
\subsection{Generic Hamiltonian with Minimal Coupling in Velocity-gauge}
I consider a generic Hamiltonian with an electric field applied using the velocity-gauge~\cite{ventura_gauge_2017, shi_geometric_2021}:
\begin{align}
\mathcal{H} = \sum_{nm, \bm{k}} \hbar \omega_{\bm{k}, n} c^\dagger_{\bm{k}, n} c_{\bm{k}, n} 
- e A^{\alpha}(\bm{q}, \omega) c^\dagger_{\bm{k}, n} \hat{v}^\alpha_{\bm{k}, \bm{k}-\bm{q}, nm} c_{\bm{k}-\bm{q}, m},
\label{eq:S1-1}
\end{align}
where  the electric field $E^{\alpha}(\bm{q}, \omega) = E^{\alpha} e^{i \bm{q} \cdot \bm{x} - i \omega t} = - i \omega A^{\alpha}(\bm{q}, \omega)$ is explicitly time and space dependent.
The reduced density matrix is defined as ~\cite{ventura_gauge_2017}, 
\begin{align}
\rho_{mn} (t) = \text{Tr} [ \hat{\rho} (t) c^\dagger_{\bm{k},n} c_{\bm{k},m} ]  = \braket{  c^\dagger_{\bm{k},n} c_{\bm{k},m} }\equiv \braket{ c^\dagger_{n} c_{m} },
\label{eq:S1-2}
\end{align}
where $\hat{\rho}(t)$ is time-dependent density operator and  the summation over the $\bm{k}$-indices is omitted. The equation of motion of the reduced density matrix is
\begin{align}
\frac{d \rho_{mn}}{d t} + i \omega_{mn} \rho_{mn} = \frac{e E^{\alpha} e^{i \bm{q} \cdot \bm{x} - i \omega t}}{i \hbar \omega} \left[ v^{\alpha}_{mp}(\bm{q}) \rho_{pn} - \rho_{mp} v^{\alpha}_{pn}(\bm{q}) \right],
\label{eq:S1-3}
\end{align}
where  $v^{\alpha}_{mp}(\bm{q}) \equiv \bra{n, \bm{k}} \hat{v}^\alpha \ket{p, \bm{k}-\bm{q}}$ is the matrix element of velocity implicitly showing the conservation of momentum. 

The reduced density matrix can be expanded in powers of the electric field as
\begin{align}
\rho_{mn}(t) = \sum_{i=0}^{\infty} \rho^{(i)}_{mn}(t).
\label{eq:S1-4}
\end{align}
The second-order density response is given by:
\begin{align}
\rho^{(2)}_{mn}(-\omega_c) &= \frac{e^2 E_a^{\alpha} E_b^{\beta} e^{i(\bm{q}_a + \bm{q}_b) \cdot \bm{x}}}{2 \hbar^2 \omega_a \omega_b (\omega_{mn} - \omega_c - 2i\eta)} \Bigg[ \frac{f_{np} v^{\beta}_{mp}(\bm{q}_b) v^{\alpha}_{pn}(\bm{q}_a)}{\omega_{pn} - \omega_a - i\eta}  - \frac{f_{pm} v^{\alpha}_{mp}(\bm{q}_a) v^{\beta}_{pn}(\bm{q}_b)}{\omega_{mp} - \omega_a - i\eta} \Bigg] + \left[ (\alpha, a) \leftrightarrow (\beta, b) \right],
\label{eq:S1-5}
\end{align}
where $f_{nm} \equiv f_n - f_m$ is the difference in Fermi distribution functions, and $\omega_{mn} \equiv \omega_m - \omega_n$ represents the energy (frequency) difference. For the second-order density response in the dc limit, the frequency and wavevector satisfy:
\begin{align}
\omega_c = 0, \quad -\omega_a = \omega_b = \omega, \quad \bm{q}_a = -\bm{q}_b = \bm{q}.
\label{eq:S1-6}
\end{align}
Consequently, Eq.~\eqref{eq:S1-5} simplifies to:
\begin{align}
\rho^{(2)}_{mn} &= \frac{e^2 E^{\alpha}_{-\omega} E^{\beta}_{\omega}}{2 \hbar^2 \omega^2} \sum_p \frac{1}{\omega_{mn} - 2i\eta} 
\Bigg[ \frac{f_{np^{-}} v^{\beta}_{mp^{-}}(\bm{q}) v^{\alpha}_{p^{-}n}(-\bm{q})}{\omega_{p^{-}n} + \omega - i\eta}  - \frac{f_{p^{+}m} v^{\alpha}_{mp^{+}}(-\bm{q}) v^{\beta}_{p^{+}n}(\bm{q})}{\omega_{mp^{+}} + \omega - i\eta} +(\alpha \leftrightarrow \beta, \bm{q} \leftrightarrow - \bm{q} ) \Bigg]
\label{eq:S1-7}
\end{align}
Here, I denote $\ket{p^{\pm}} \equiv \ket{p, \bm{k} \pm \bm{q}}$ to explicitly show the momentum transfer of states.


\subsection{Second-order Photovoltaic Effects}
The second-order photovoltaic response is derived from the contraction of the reduced density matrix with the velocity matrix~\cite{ventura_gauge_2017}:
\begin{align}
J^{(2),\gamma} = e\sum_{n,m,\bm{k}} v^{\gamma}_{nm} \rho^{(2)}_{mn},
\label{eq:S1-8}
\end{align}
typically divided into diagonal (injection) and off-diagonal (shift) currents:
\begin{align}
J^{(2),\gamma}_{\text{inj}} = e\sum_{n,\bm{k}} v^{\gamma}_{nn} \rho^{(2)}_{nn}, \quad
J^{(2),\gamma}_{\text{shift}} = e\sum_{\tilde{n} \neq \tilde{m},\bm{k}} v^{\gamma}_{nm} \rho^{(2)}_{mn},
\label{eq:S1-9}
\end{align}
where $\tilde{n}$ denotes the energy-degenerate subspace. Note that for degenerate bands ($\omega_{n} = \omega_{m}$), the off-diagonal matrix element of the velocity within the degenerate subspace vanishes:
\begin{align}
v^{\gamma}_{nm} &= \frac{1}{\hbar}\bra{n} \partial_{k^\gamma} H\ket{m} = v_n \delta_{nm} + (\omega_m - \omega_n) \braket{n | \partial_{k_\gamma} m}.
\label{eq:S1-10}
\end{align}

\subsection{Formula for Injection Current of the PDE}
By symmetrizing the external photon wavevector $\bm{q}$ in the density response $\rho^{(2)}_{mn}$, the injection current can be expressed as
\begin{align}
J^{\gamma}_{\text{inj.}} &= \frac{e^3 E^{\alpha}_{-\omega} E_{\omega}^{\beta}}{4i \eta \hbar^2 \omega^2} \sum_{n,m,\bm{k}} (v^{\gamma}_{n^{+}} - v^{\gamma}_{m^{-}}) f_{n^{+}m^{-}} v^{\beta}_{n^{+} m^{-}} v^{\alpha}_{m^{-}n^{+}} \left( \frac{1}{\omega_{n^{+}m^{-}} - \omega + i \eta}- \text{c.c.} \right) \notag \\
&= - \frac{\pi e^3 E^{\alpha}_{-\omega} E_{\omega}^{\beta}}{2\eta \hbar^2 \omega^2} \sum_{n,m,\bm{k}} f_{n^{+}m^{-}} (v^{\gamma}_{n^{+}} - v^{\gamma}_{m^{-}}) v^{\beta}_{n^{+}m^{-}} v^{\alpha}_{m^{-}n^{+}} \delta(\omega_{n^{+}m^{-}} - \omega),
\label{eq:S1-11}
\end{align}
where I define $\ket{n^+} = \ket{n, \bm{k} + \bm{q}/2}$ and $\ket{m^-} = \ket{m, \bm{k} - \bm{q}/2}$. The conductivity tensor for injection current [Eq. (3) in the main text] becomes:
\begin{align}
\sigma^{\gamma;\alpha \beta}_{\text{inj.}} (\bm{q}) = - \frac{e^3}{2 \eta \hbar^2 \omega^2} \sum_{n,m,\bm{k}} f_{nm}(\bm{q}) \Delta^{\gamma}_{nm}(\bm{q}) v^{\beta}_{nm}(\bm{q}) v^{\alpha}_{mn}(-\bm{q}) \delta(\omega_{nm}(\bm{q}) - \omega),
\label{eq:S1-12}
\end{align}
where I show the momentum transfer as a function of $\bm{q}$ and omit the superscripts ($\pm$) for clarity. $\Delta^{\gamma}_{nm}(\bm{q}) = v^\gamma_{n,\bm{k} + \bm{q}/2} - v^\gamma_{m,\bm{k} - \bm{q}/2}$ represents the difference in group velocity.

\subsection{Formula for Shift Current of the PDE}
In similar way, the shift current is expressed as
\begin{align}
J^{\gamma}_{\text{shift}} &= \frac{e^3 E^{\alpha}_{-\omega} E_{\omega}^{\beta}}{2 \hbar^2 \omega^2} \sum_{\tilde{n} \neq \tilde{m},p,\bm{k}} \frac{v^\gamma_{nm}}{\omega_{mn}} \left[ \frac{f_{n p^{-}} v^{\beta}_{m p^{-}} v^{\alpha}_{p^{-}n}}{\omega_{p^{-}n} + \omega - i \eta} - \frac{f_{p^{+} m} v^{\alpha}_{p^{+}m} v^{\beta}_{p^{+}n}}{\omega_{mp^{+}} + \omega - i \eta} + (\alpha \leftrightarrow \beta, \bm{q} \leftrightarrow - \bm{q} )  \right].
\label{eq:S1-13}
\end{align}
 Employing the identity:
\begin{align}
\sum_{\tilde{n} \neq \tilde{m}} \ket{n} \frac{v^{\gamma}_{nm}}{\omega_{mn}} &= \ket{\partial_{\gamma} m} + \sum_{\tilde{l} = \tilde{m}} i A^\gamma_{lm} \ket{l}.
\label{eq:S1-14}
\end{align}
the equation reduces to
\begin{align}
J^{\gamma}_{\text{shift}} 
&= \frac{e^3 E^{\alpha}_{-\omega} E_{\omega}^{\beta}}{2 \hbar^2 \omega^2} \sum_{n,m,\bm{k}} f_{n^{+}m^{-}} \Bigg\{ 
\frac{\left[ \partial_{\gamma} v^{\beta}_{n^{+} m^{-}} - i \left( \sum\limits_{\tilde{l} = \tilde{n}} A^\gamma_{n^{+}l^{+}} v^{\beta}_{l^{+} m^{-}} - \sum\limits_{\tilde{l} = \tilde{m}} v^{\beta}_{n^{+} l^{-}} A^\gamma_{l^{-}m^{-}} \right) - w^{\gamma\beta}_{n^+ m^-} \right] v^{\alpha}_{m^{-}n^{+}}}{\omega_{n^{+}m^{-}} - \omega + i \eta} \notag \\
& + \frac{v^{\beta}_{n^{+}m^{-}} \left[ \partial_{\gamma} v^{\alpha}_{m^{-} n^{+}} - i \left( \sum\limits_{\tilde{l} = \tilde{m}} A^\gamma_{m^{-}l^{-}} v^{\alpha}_{l^{-} n^{+}} - \sum\limits_{\tilde{l} = \tilde{n}} v^{\alpha}_{m^{-} l^{+}} A^\gamma_{l^{+}n^{+}} \right) - w^{\gamma\alpha}_{m^-n^+} \right]}{\omega_{n^{+}m^{-}} - \omega - i \eta} \Bigg\},
\label{eq:1-15}
\end{align}
where \( w^{\gamma\beta}_{n^+ m^-} = \bra{n^+} \partial_\gamma \hat{v}^\beta \ket{m^-} = \frac{1}{\hbar} \bra{n^+} \partial_\gamma \partial_\beta \hat{H} \ket{m^-} \) is the diamagnetic velocity, which vanishes with linear band dispersions. Neglecting the diamagnetic term, the conductivity tensor for shift current [Eq. (4) in the main text] simplifies to:
\begin{align}
\sigma^{\gamma; \alpha \beta}_{\text{shift}} (\bm{q}) &= \frac{e^3}{2 \hbar^2 \omega^2} \sum_{n,m,\bm{k}} f_{nm} (\bm{q}) \left[ 
 \frac{v^{\beta}_{nm} (\bm{q}) \mathcal{D}_\gamma v^{\alpha}_{mn} (-\bm{q})}{\omega_{nm} (\bm{q}) - \omega - i \eta} + \frac{\mathcal{D}_\gamma v^{\beta}_{nm} (\bm{q}) v^{\alpha}_{mn} (-\bm{q})}{\omega_{nm} (\bm{q}) - \omega + i \eta}  \right] .
\label{eq:S1-16}
\end{align}

\section{Symmetry operations in the Dirac Hamiltonian}\label{Sup:B}

I recall the Dirac Hamiltonian in the main text [Eq.~(6)]
\begin{align}
H = \Delta \gamma_0 + i  \hbar v  k_i \gamma_0 \gamma_i .
\label{eq:S2-1}
\end{align}
with eigenenergies and eigenstates:
\begin{align}
\label{eq:S2-2}
\omega_{\tilde{c}} &= \omega_k , \omega_{\tilde{v}}= - \omega_k ,\\
u_{c,1}&= \sqrt{ \frac{ \omega_k +\tilde{\Delta}}{ 2 \omega_k  } } 
\left(  -1 , 0, \frac{i v k_3}{  \omega_k +\tilde{\Delta} } , \frac{i v k_1 - v k_2}{  \omega_k +\tilde{\Delta} } \right) ,\notag \\
u_{c,2}&= \sqrt{ \frac{ \omega_k +\tilde{\Delta}}{ 2 \omega_k  } } 
\left(  0 , -1, \frac{iv k_1 +v k_2}{  \omega_k +\tilde{\Delta} } , \frac{-i v k_3}{  \omega_k +\tilde{\Delta} }  \right) ,\notag\\
u_{v,1}&=\sqrt{ \frac{ \omega_k +\tilde{\Delta}}{ 2 \omega_k  } } 
\left(  \frac{i v k_3}{  \omega_k +\tilde{\Delta} } , \frac{i v k_1 - v k+2}{  \omega_k +\tilde{\Delta} } ,  -1 , 0 \right) ,\notag \\
u_{v,2}&= \sqrt{ \frac{ \omega_k +\tilde{\Delta}}{ 2 \omega_k } } 
\left(  \frac{i v k_1 + v k_2}{ \omega_k +\tilde{\Delta} } , \frac{-i v k_3}{ \omega_k +\tilde{\Delta} }  , 0 , -1\right) ,
\label{eq:S2-3}
\end{align}
where $ \omega_k = \sqrt{ v^2 k^2  +\tilde{\Delta}^2}$ with normalized half band gap $\tilde{\Delta} \equiv \Delta/\hbar$. $\tilde{c}, \tilde{v}$ denotes the conduction (valance) band subspace.

\subsection{Charge Conjugation in the Dirac Hamiltonian}
The Dirac Hamiltonian  respects charge conjugation ($\mathcal{C}$) symmetry, defined by:
\begin{align}
\mathcal{C} = i \gamma_2 K ,
\label{eq:S2-4}
\end{align}
where $K$ is complex conjugate. $\mathcal{C}$ is anti-unitary operator and transforms a conduction state $u_c$ into a valance state $u_v$:
\begin{align}
\mathcal{C} u_{c,\sigma} (\bm{k}) =   u_{v,\bar{\sigma}} (\bm{k}) (i \sigma_2)_{ \bar{\sigma} \sigma} ,
\label{eq:S2-5}
\end{align}
where $\sigma_2$ is the Pauli matrix and $ \sigma,\bar{\sigma}$ denotes the spin indices.
The Hamiltonian and velocity operator transform under charge conjugation as
\begin{align}
\mathcal{C} \hat{H} \mathcal{C}^{-1} = - \hat{H}, \quad \mathcal{C} \hat{v}_i \mathcal{C}^{-1} = -\hat{v}_i  ,
\label{eq:S2-6}
\end{align}
Due to the charge conjugation (particle-hole) symmetry, the interband matrix element of velocity and covariant derivative on velocity satisfy the following relations:
\begin{align}
\label{eq:S2-7}
v^\beta_{c,\sigma^+ ; v,\tau^-} &= \bra{  u_{c,\sigma} (\bm{k} + \bm{q}/2) } \hat{v}^\beta \ket{u_{v,\tau}  (\bm{k} - \bm{q}/2)} \notag \\
&=  \bra{  u_{c,\sigma} (\bm{k} + \bm{q}/2) } \mathcal{C}^{-1} \left( - \hat{v}^\beta \right)\mathcal{C} \ket{u_{v,\tau}  (\bm{k} - \bm{q}/2)} \notag \\
&= -  \left( \bra{ \mathcal{C} u_{c,\sigma} (\bm{k} + \bm{q}/2) }   \hat{v}^\beta  \ket{ \mathcal{C} u_{v,\tau}  (\bm{k} - \bm{q}/2)} \right)^* \notag \\
&= -  \left( \bra{  u_{v,\bar{\sigma}} (\bm{k} + \bm{q}/2) }   \hat{v}^\beta  \ket{  u_{c,\bar{\tau}}  (\bm{k} - \bm{q}/2)} \right)^* (i \sigma)_{  \bar{\sigma}  \sigma} (i \sigma)_{ \bar{\tau} \tau} \notag \\
&= -   \bra{  u_{c,\bar{\tau}} (\bm{k} - \bm{q}/2) } \hat{v}^\beta  \ket{  u_{v,\bar{\sigma}}  (\bm{k} + \bm{q}/2)} (i \sigma_2)_{   \bar{\sigma} \sigma}(i \sigma_2))_{ \bar{\tau} \tau} \notag \\
&= - v^\beta_{c,\bar{\tau}^- ; v,\bar{\sigma}^+}  (i \sigma_2)_{ \bar{\sigma} \sigma} (i \sigma_2)_{ \bar{\tau} \tau} , \\
\mathcal{D}_\gamma v^\beta_{c,\sigma^+; v,\tau^-} &= \partial_{\gamma} v^\beta_{c,\sigma^+; v,\tau^-} - i\left( \sum\limits_{\rho } A^\gamma_{c,\sigma^+;c,\rho^+} v^\beta_{c,\rho^+; v,\tau^-} - v^\beta_{c,\sigma^+; v,\rho^-} A^\gamma_{v,\rho^-;v,\tau^-} \right) \notag \\
&= -  \partial_{\gamma} v^\beta_{c,\bar{\tau}^-; v,\bar{\sigma}^+} (i \sigma_2)_{\bar{\sigma} \sigma}  (i \sigma_2)_{\bar{\tau} \tau} - i\Big( \sum\limits_{\rho }   v^\beta_{c,\bar{\tau}^-; v, \bar{\rho}_1^+} A^\gamma_{v,\bar{\rho}_2^+;v,\bar{\sigma}^+} (i \sigma_2)_{\bar{\sigma} \sigma}  (i \sigma_2)_{\bar{\rho}_1 \rho} (i \sigma_2)_{\bar{\rho}_2 \rho} (i \sigma_2)_{\bar{\tau} \tau} \notag \\
& - A^\gamma_{c,\bar{\tau}^-;c,\bar{\rho}_1^-} v^\beta_{c,\bar{\rho}_2^-; v,\bar{\sigma}^+} (i \sigma_2)_{\bar{\sigma} \sigma}  (i \sigma_2)_{\bar{\rho}_1 \rho} (i \sigma_2)_{\bar{\rho}_2 \rho} (i \sigma_2)_{\bar{\tau} \tau} \Big)  \notag \\
&= - (i \sigma_2)_{\bar{\sigma} \sigma}  (i \sigma_2)_{\bar{\tau} \tau} \left[ \partial_{\gamma} v^\beta_{c,\bar{\tau}^-; v,\bar{\sigma}^+} - i\left(\sum\limits_{\rho }  A^\gamma_{c,\bar{\tau}^-;c,\bar{\rho}^-} v^\beta_{c,\bar{\rho}^-; v,\bar{\sigma}^+} -   v^\beta_{c,\bar{\tau}^-; v, \bar{\rho}^+} A^\gamma_{v,\bar{\rho}^+;v,\bar{\sigma}^+}    \right) \right] \notag \\
&=  - \mathcal{D}_\gamma v^\beta_{c,\bar{\tau}^-; v,\bar{\sigma}^+}    (i \sigma_2)_{\bar{\sigma} \sigma}  (i \sigma_2)_{\bar{\tau} \tau} ,
\label{eq:S2-8}
\end{align}
where I denote $\bar{\sigma} = -\sigma$ for spin-$\frac{1}{2}$ fermions. Note that the intraband (within the degenerate subspace) Berry connection follows the relation:
\begin{align}
A^\gamma_{c,\sigma^+;c,\tau^+} &=   i \bra{ u_{c,\sigma} (+ ) } \mathcal{C}^{-1} \partial_{\gamma} \mathcal{C} \ket{u_{c,\tau} ( + )} \notag \\
&=   i \left( \bra{ \mathcal{C} u_{c,\sigma} (+ ) }  \partial_{\gamma}  \ket{\mathcal{C} u_{c,\tau} ( + )} \right)^* \notag \\
&=   i \left( \bra{  u_{v, \bar{\sigma}} (+ ) }  \partial_{\gamma}  \ket{ u_{v,\bar{\tau}} ( + )} \right)^* (i \sigma_2)_{ \bar{\sigma} \sigma} (i \sigma_2)_{ \bar{\tau} \tau} \notag \\
&=   i \bra{  u_{v, \bar{\tau}} (+ ) }  \partial_{\gamma}  \ket{ u_{v,\bar{\sigma}} ( + )} (i \sigma_2)_{ \bar{\sigma}   \sigma} (i \sigma_2)_{ \bar{\tau} \tau} \notag \\
&=   A^\gamma_{v,\bar{\tau}^+;v,\bar{\sigma}^+}  (i \sigma_2)_{ \bar{\sigma} \sigma} (i \sigma_2)_{ \bar{\tau} \tau} .
\label{eq:S2-9}
\end{align}

\subsection{ $\mathcal{P},\mathcal{T}$-inversion in the Dirac Hamiltonian}
The spatial $(\mathcal{P})$ and time $(\mathcal{T})$ inversion operator  in the Dirac Hamiltonian are
\begin{align}
\mathcal{T} = \frac{1}{2} \gamma_0 \left[ \gamma_3 , \gamma_1 \right] K,  \quad \mathcal{P}  =\gamma_0, 
\label{eq:S2-10}
\end{align}
The $\mathcal{T}$-operator inverts both spin and momentum of eigenstates:
\begin{align}
\mathcal{T} u_{c,\sigma} (\bm{k}) =  (-1)^{c,v}  u_{c,\bar{\sigma}} (-\bm{k}) (i \sigma_2)_{ \bar{\sigma} \sigma}, 
\label{eq:S2-11}
\end{align}
where $c=0,v=1$, indicating even and odd parity for conduction and valance bands, respectively.

The $\mathcal{P}$-operator only inverts the momentum:
\begin{align}
\mathcal{P} u_{c,\sigma} (\bm{k}) = (-1)^{c,v}  u_{c,\sigma}  (-\bm{k}) .
\label{eq:S2-12}
\end{align}

Due to the disparate parity between the conduction and valance states, the intraband and interband  matrix elements of velocity and Berry connection follow different transformation rules. Under the $\mathcal{T}$-inversion, the intraband matrix elements of Berry connection (at $\bm{q}=0$) satisfy the following relation:
\begin{align}
A^\gamma_{c,\sigma;c,\tau} (\bm{k}) &= A^\gamma_{c,\bar{\tau};c,\bar{\sigma}} (-\bm{k}) (i \sigma_2)_{ \bar{\sigma} \sigma} (i \sigma_2)_{ \bar{\tau} \tau}.
\label{eq:S2-13}
\end{align}

For the interband  matrix elements of Berry connection and covariant derivative on Berry connection, the transformation under $\mathcal{T}$-inversion is given by: 
\begin{align}
\label{eq:S2-13}
A^\gamma_{c,\sigma;v,\tau} (\bm{k}) &= -A^\gamma_{v,\bar{\tau};c,\bar{\sigma}} (-\bm{k}) (i \sigma_2)_{ \bar{\sigma} \sigma} (i \sigma_2)_{ \bar{\tau} \tau},\\
\mathcal{D}_\gamma A^{\beta}_{c,\sigma; v, \tau} ( \bm{k}) 
&=\mathcal{D}_{\bar{\gamma}} A^{\beta}_{v,\bar{\tau}; c, \bar{\sigma}} ( - \bm{k})    (i \sigma_2)_{ \bar{\tau} \tau}  (i \sigma_2)_{ \bar{\sigma} \sigma},
\label{eq:S2-14}
\end{align}
where I use identity: $\partial_{\bar{\gamma} } \equiv \partial_{k^{\bar{\gamma}} }= \partial_{- k^{\gamma} } = -  \partial_{\gamma }$.

Thus, the quantum metric connection has relation
\begin{align}
C^{\beta \gamma \alpha}_{\tilde{c},\tilde{v}} (\bm{k}) &= \sum_{\sigma,\tau}  A^{\beta}_{c,\sigma; v, \tau} (\bm{k})\mathcal{D}_\gamma  A^{\alpha}_{ v, \tau;c,\sigma}(\bm{k}) = - \sum_{\sigma,\tau}  A^{\beta}_{c,\sigma; v, \tau} (-\bm{k})\mathcal{D}_\gamma  A^{\alpha}_{ v, \tau;c,\sigma}(-\bm{k}) = -C^{\beta \gamma \alpha}_{\tilde{v},\tilde{c}} (-\bm{k}). 
\label{eq:S2-15}
\end{align}

Under the $\mathcal{P}$-inversion, the intraband matrix elements of  Berry connection obey the following relation:
\begin{align}
A^\gamma_{c,\sigma;c,\tau} (\bm{k}) &=-  A^\gamma_{c,\sigma;c,\tau}  (-\bm{k}).
\label{eq:S2-16}
\end{align}

For the interband matrix elements of Berry connection and covariant derivative on Berry connection the relations become:
\begin{align}
\label{eq:2-17}
A^\beta_{c,\sigma ; v,\tau}  (\bm{k})  &= A^\beta_{c, \sigma ; v, \tau} (- \bm{k}) , \\
\mathcal{D}_\gamma A^{\beta}_{c,\sigma; v, \tau} ( \bm{k}) &= -\mathcal{D}_{\bar{\gamma}} A^{\beta}_{c,\sigma; v, \tau} ( - \bm{k}).
\label{eq:2-18}
\end{align}
Thus, the quantum metric connection satisfies the following relation:
\begin{align}
C^{\beta \gamma \alpha}_{\tilde{c},\tilde{v}} (\bm{k}) &= \sum_{\sigma,\tau}  A^{\beta}_{c,\sigma; v, \tau} (\bm{k}) \mathcal{D}_\gamma  A^{\alpha}_{ v, \tau;c,\sigma}(\bm{k}) =   - \sum_{\sigma,\tau}  A^{\beta}_{c,\sigma; v, \tau} (-\bm{k}) \mathcal{D}_\gamma  A^{\alpha}_{ v, \tau;c,\sigma}(-\bm{k}) = - C^{\beta \gamma \alpha}_{\tilde{c},\tilde{v}} (-\bm{k})  . 
\label{eq:2-19}
\end{align}

\section{Photon Drag Effect in the Dirac Hamiltonian }\label{Sup:C}

\subsection{Intraband Effect: Injection Photon Drag Effect}
I first consider the injection current of the first-order PDE ($\propto q^\tau$). The $\bm{q}$-relevant quantities in Eq.~\eqref{eq:S1-12} are: 1. Fermi distribution function $f_{nm} (\bm{q})$; 2. $\delta$-function $\delta \left(    \omega_{nm} (\bm{q}) -  \omega   \right) $; 3.  group velocity $\Delta^{\gamma}_{nm} (\bm{q})$, and 4. product of velocity matrix elements $v^{\beta}_{n m}(\bm{q})   v^{\alpha}_{mn}(-\bm{q})$. The first three quantities can be expanded in powers of photon wavevector $\bm{q}$ as follows:

1. Fermi distribution expansion:
\begin{align}
f_{nm} (\bm{q}) 
&= f_n (\bm{k} ) - f_m (\bm{k}) + \frac{\hbar q^\tau}{2} \left[ f'_n (\bm{k}) v^\tau_n (\bm{k}) + f'_m (\bm{k}) v^\tau_m (\bm{k}) \right] + \mathcal{O} (q^2) \\
&= f_{nm} + \frac{\hbar q^\tau}{2} d^\tau_{f} + \mathcal{O} (q^2).
\label{eq:S3-1}
\end{align}

 2. $\delta$-function expansion:
\begin{align}
2\pi i  \delta \left(    \omega_{nm} (\bm{q}) -  \omega   \right)  &= \left(  \frac{1 }{    \omega_{nm} (\bm{q}) -  \omega - i \eta } -   \frac{  1 }{    \omega_{nm}  (\bm{q}) -  \omega  + i \eta}   \right) \notag\\
  & =  2\pi i  \delta \left(    \omega_{nm}  -  \omega   \right)
  + 2\pi q^\tau \left( v^\tau_{n} + v^\tau_{m} \right)  \text{Im} \frac{ 1 }{   ( \omega_{nm}  -  \omega - i \eta )^2 }   + \mathcal{O} (q^2) .
\label{eq:S3-2}
\end{align}

3. Group velocity expansion:
\begin{align}
 \Delta^{\gamma}_{nm} (\bm{q}) &= \Delta^{\gamma}_{nm}  + \frac{q^\tau}{2}  \partial_{\tau}\left( v^\gamma_{n} + v^\gamma_{m}   \right) + \mathcal{O} (q^2).
\label{eq:S3-3}
\end{align}

Depending on the Fermi distribution function, I divide the $\bm{q}$-expansion of Eq.~\eqref{eq:S1-12}  into \textit{Fermi surface} and  \textit{Fermi sea} contributions. The  \textit{Fermi surface} contribution results from the $\bm{q}$-expansion of Fermi distribution [Eq.~(\ref{eq:S3-1})]:
\begin{align}
\chi^{\gamma \tau ;\alpha \beta}_{\text{inj.}} 
&= -  \frac{ \pi e^3  }{ 4 \hbar \omega^2 \eta } \sum_{ \sigma, \tau, \bm{k} }  d^\tau_{f}   v^{\beta}_{c,  \sigma ; v, \tau}   v^{\alpha}_{v,\tau;  c,\sigma}  \Delta^{\gamma}_{cv}   \delta (  \omega_{cv} -  \omega  )  .
\label{eq:S3-4}
\end{align}
where the summation within degenerate subspace is excluded, since $\Delta^{\gamma}_{n m}=0$ for $\tilde{m} = \tilde{n}$. 

The  \textit{Fermi sea} contribution contains the Fermi-distribution difference $f_{nm}$ ($\bm{q}=0$) which requires $\tilde{n} \neq \tilde{m}$. Under particle-hole symmetry, the opposite group velocity of valance and conduction bands cancel the group velocity [Eq.~\eqref{eq:S3-3}] and $\delta$-function [Eq.~\eqref{eq:S3-2}] terms, leaving only the product of the interband velocity matrix elements: 
\begin{align}
\chi^{\gamma \tau ;\alpha \beta}_{\text{inj.}} &=  - \frac{e^3   }{  2 \eta \hbar^2 \omega^2 } \sum_{\tilde{n} \neq \tilde{m}, \bm{k} }  f_{nm}    \partial_{\tau} \Big[ v^{\beta}_{n m}(\bm{q})   v^{\alpha}_{mn}(-\bm{q})  \Big] |_{\bm{q}=0} \Delta^{\gamma}_{nm}  \delta \left(     \omega_{nm}  -  \omega    \right) .
\label{eq:S3-5}
\end{align}
However, due to $\mathcal{C}$-symmetry of the Dirac Hamiltonian [Eq.~(\ref{eq:S2-7})], the product of interband velocity matrix elements satisfies:
\begin{align}
v^\beta_{c,\sigma^+ ; v,\tau^-} v^\alpha_{v,\tau^- ; c,\sigma^+} &=   v^\beta_{c,\bar{\tau}^- ; v,\bar{\sigma}^+} v^\alpha_{v,\bar{\sigma}^+ ; c,\bar{\tau}^-}   \notag \\
\partial_{q^\tau} \left( v^\beta_{c,\sigma ; v,\tau} v^\alpha_{v,\tau ; c,\sigma} \right) &= -\partial_{q^\tau} \left( v^\beta_{c,\bar{\tau} ; v,\bar{\sigma}} v^\alpha_{v,\bar{\sigma} ; c,\bar{\tau}} \right),
\label{eq:S3-6}
\end{align}
which vanishes  upon summation over degenerate subspace $\tilde{c},\tilde{v}$. Thus, only the \textit{Fermi surface} term contributes to the injection current of the PDE. 

The injection current [Eq.~\eqref{eq:S3-4}] can be expressed in terms of the quantum metric tensor
\begin{align}
Q^{\beta \alpha}_{\tilde{c} ; \tilde{v}} =  \sum_{ \sigma, \tau}  A^{\beta}_{c,  \sigma ; v, \tau}   A^{\alpha}_{v,\tau;  c,\sigma} 
=  \sum_{ \sigma, \tau} \omega^2_{cv}   v^{\beta}_{c,  \sigma ; v, \tau}   v^{\alpha}_{v,\tau;  c,\sigma} = g^{\beta \alpha}_{\tilde{c} ; \tilde{v}} -  \frac{i}{ 2 } \Omega^{\beta \alpha}_{\tilde{c} ; \tilde{v}}.
\label{eq:S3-7}
\end{align}
where $ g^{\beta \alpha}_{\tilde{c} ; \tilde{v}}$ is the quantum metric and $\Omega^{\beta \alpha}_{\tilde{c} ; \tilde{v}}$ is the Berry curvature.

The responses to linearly polarized (LP) and circularly polarized (CP) lights are obtained by symmetrizing and antisymmetrizing the indices $\alpha$,$\beta$, corresponding to the quantum metric and Berry curvature, respectively:
\begin{align}
\label{eq:S3-8}
\chi^{\gamma \tau ;\alpha \beta}_{L,\text{inj.}}
&= -  \frac{ \pi e^3   }{ 4 \hbar \omega^2 \eta } \sum_{  \bm{k} }  \left( f'_c  v^\tau_c + f'_v v^\tau_v \right) \Delta^{\gamma}_{cv} \omega^2_{cv}   \tilde{g}^{ \beta \alpha}_{\tilde{c} ; \tilde{v}}    \delta (  \omega_{cv} -  \omega  ) , \\
\chi^{\gamma \tau ; \sigma }_{C,\text{inj.}} 
&= - \epsilon_{\alpha \beta \sigma}  \frac{ \pi e^3  }{ 8  \hbar \omega^2 \eta } \sum_{ \bm{k} }  \left( f'_c  v^\tau_c + f'_v v^\tau_v \right) \Delta^{\gamma}_{cv} \omega^2_{cv}  \tilde{\Omega}^{\beta \alpha}_{\tilde{c} ; \tilde{v}}   \delta (  \omega_{cv} -  \omega  ).
\label{eq:S3-9}
\end{align}

For the Dirac Hamiltonian, the quantum metric and the Berry curvature take the following form:
\begin{align}
g^{\beta \alpha}_{\tilde{c} ; \tilde{v}} = \frac{v_F^2}{2} \left(\frac{ \delta_{\beta \alpha}}{\omega_{k}^2} - \frac{v_F^2 k^\beta k^\alpha}{  \omega_{k}^4} \right) ,\quad \Omega^{\beta \alpha}_{\tilde{c} ; \tilde{v}} = 0 .
\label{eq:S3-10}
\end{align}

Thus, for the Dirac Hamiltonian, only LP light contributes to the injection current response:
\begin{align}
\chi^{\gamma \tau ;\alpha \beta}_{L,\text{inj.}} 
&= -  \frac{ 2 \pi e^3  v^2 }{  \hbar \omega^2 \eta } \sum_{  \bm{k} }  \left( f'_c  - f'_v \right)  k^\tau k^\gamma  \left( \frac{ \delta_{\beta \alpha}}{ \omega_{k}^2} - \frac{ v^2 k^\beta k^\alpha}{  \omega_{k}^4} \right)   \delta ( 2 \omega_k -  \omega  ) .
\label{eq:S3-11}
\end{align}

\subsection{Interband effect: Shift Photon Drag Effect}
The shift current of the PDE can be obtained in a similar way to the injection current. The Fermi surface contribution to the shift PDE reads:
\begin{align}
\chi^{\gamma \tau ; \alpha \beta}_{\text{shift}}  &= \frac{e^3  }{ 4 \hbar \omega^2}   \sum_{\sigma, \tau,\bm{k}}  d^\tau_{f} \left[ \frac{ v^{\beta}_{c,\sigma; v, \tau}  \mathcal{D}_\gamma v^{\alpha}_{v, \tau;c,\sigma}   }{ \omega_{cv} -\omega - i \eta }  + \frac{ \mathcal{D}_\gamma v^{\beta}_{c,\sigma; v, \tau}  v^{\alpha}_{ v, \tau;c,\sigma}  }{ \omega_{cv} -\omega + i \eta} \right] .
\label{eq:S3-12}
\end{align}
where I neglect the summation within degenerate subspace ($\tilde{n}= \tilde{m}$), which causes infrared divergence.

The  \textit{Fermi sea} contribution arises from the $\bm{q}$-expansion of the product of interband matrix elements velocity and its covariant derivatives: 
\begin{align}
\chi^{\gamma \tau; \alpha \beta}_{\text{shift}}  &= \frac{e^3   }{ 2 \hbar^2 \omega^2}   \sum_{\tilde{n} \neq \tilde{m},\bm{k}}  f_{nm} \left\{   \frac{ \partial_{\tau} \left[ v^{\beta}_{nm} (\bm{q})  \mathcal{D}_\gamma v^{\alpha}_{mn} (-\bm{q}) \right]  }{ \omega_{nm} -\omega - i \eta }  + \frac{ \partial_{\tau} \left[ \mathcal{D}_\gamma v^{\beta}_{nm} (\bm{q}) v^{\alpha}_{mn} (-\bm{q}) \right]  }{ \omega_{nm} -\omega + i \eta}  \right\} .
\label{eq:S3-13}
\end{align}
which vanishes after summing the degenerate subspace, due to the constraint of $\mathcal{C}$-symmetry of the Dirac Hamiltonian  [Eq.~(\ref{eq:S2-7}\&\ref{eq:S2-8})]:
\begin{align}
 v^\beta_{c,\sigma^+; v,\tau^-}  \mathcal{D}_\gamma v^\alpha_{v,\tau^- ; c,\sigma^+} &=     v^\beta_{c,\bar{\tau}^-; v,\bar{\sigma}^+}   \mathcal{D}_\gamma v^\alpha_{v,\bar{\sigma}^+ ; c,\bar{\tau}^-} \notag \\
 \partial_{q^\tau} \left( \mathcal{D}_\gamma v^\beta_{c,\sigma^+; v,\tau^-}  v^\alpha_{v,\tau^- ; c,\sigma^+} \right) &=  -\partial_{q^\tau} \left(   \mathcal{D}_\gamma v^\beta_{c,\bar{\tau}^-; v,\bar{\sigma}^+}  v^\alpha_{v,\bar{\sigma}^+ ; c,\bar{\tau}^-} \right).
\label{eq:S3-14}
\end{align}
Similarly, shift photon-drag current only has  \textit{Fermi surface}  contribution [Eq.~(\ref{eq:S3-11})] in the Dirac electrons. 

\subsection{Quantum Metric Connection}

To discuss the geometric properties of the shift PDE, I define the quantum metric connection following Refs.~\cite{ahn_low-frequency_2020,ahn_riemannian_2022}:
\begin{align}
C^{\beta \gamma \alpha}_{\tilde{c},\tilde{v}} = \sum_{\sigma,\tau} A^{\beta}_{c,\sigma; v, \tau}  \mathcal{D}_\gamma  A^{\alpha}_{ v, \tau;c,\sigma},
\label{eq:S3-15}
\end{align}
Note that the interband Berry connection  is equivalent to interband matrix elements of position operator, $A^{\beta}_{c,\sigma; v, \tau} = r^{\beta}_{c,\sigma; v, \tau}$ which serves as the tangent basis vector of Riemannian geometry. Additionally, the covariant derivative on right is complex conjugate of $C^{\beta\alpha;\gamma}_{\tilde{c},\tilde{v}} $ with indices $\alpha \leftrightarrow \beta$  interchanged:
\begin{align}
 \sum_{\sigma,\tau}  \mathcal{D}_\gamma A^{\beta}_{c,\sigma; v, \tau}  A^{\alpha}_{ v, \tau;c,\sigma} ,
 &= \left( \sum_{\sigma,\tau}   A^{\alpha}_{c,\sigma;  v, \tau}   \mathcal{D}_\gamma A^{\beta}_{ v, \tau; c,\sigma}   \right)^* =    \left( C^{ \alpha \gamma \beta }_{\tilde{c},\tilde{v}}  \right)^* .
\label{eq:S3-16}
\end{align}

The covariant derivative on interband Berry connection and covariant derivative on velocity matrix elements  satisfy the following identity:
\begin{align}
\mathcal{D}_\mu A^\nu_{ab} &= -i \partial_{\mu} \left( \frac{ v^\nu_{ab}}{\omega_{ab} } \right) -  \left( \sum_{\tilde{c} = \tilde{a} } A^\mu_{ac} \frac{ v^\nu_{cb} }{ \omega_{ab}} - \sum_{\tilde{c} = \tilde{a} } \frac{  v^\nu_{ac} }{ \omega_{ab}}A^\mu_{cb}   \right) = \frac{-i}{ \omega_{ab}  } \mathcal{D}_\mu v^\nu_{ab} + \frac{ i v^\nu_{ab} (v^\mu_a - v^\mu_b)}{\omega^2_{ab} }  .
\label{eq:S3-17}
\end{align}
Employ this identity, the product of interband matrix elements of velocity and its covariant derivative can be rewritten as
\begin{align}
 \sum\limits_{\sigma,\tau}  v^{\beta}_{c,\sigma; v, \tau}   D_\gamma v^{\alpha}_{ v, \tau;c,\sigma} &= \omega^2_{cv} \left[ C^{\beta \gamma \alpha}_{\tilde{c},\tilde{v}}  +  \frac{ \Delta^\gamma_{cv} }{\omega_{cv} } Q^{\beta\alpha}_{\tilde{c},\tilde{v}} \right], \notag \\
  \sum\limits_{\sigma,\tau}  D_\gamma  v^{\beta}_{c,\sigma; v, \tau}   v^{\alpha}_{ v, \tau;c,\sigma} &= \omega^2_{cv} \left[( C^{\alpha \gamma \beta}_{\tilde{c},\tilde{v}})^* + \frac{ \Delta^\gamma_{cv} }{\omega_{cv} } (Q^{\alpha \beta}_{\tilde{c},\tilde{v}} )^*\right] .
\label{eq:S3-18}
\end{align}

The LP light response is obtained by symmetrizing the indices $\alpha,\beta$:
\begin{align}
\chi^{\gamma \tau ; \alpha \beta}_{L,\text{shift}}  &= \frac{e^3  }{ 8 \hbar \omega^2}   \sum_{\bm{k}}  d^\tau_{f} \omega^2_{cv} \left( \frac{ C^{\beta \gamma \alpha}_{\tilde{c},\tilde{v}} +   \frac{ \Delta^\gamma_{cv} }{\omega_{cv} } Q^{\beta\alpha}_{\tilde{c},\tilde{v}} + C^{\alpha \gamma \beta}_{\tilde{c},\tilde{v}} +   \frac{ \Delta^\gamma_{cv} }{\omega_{cv} } Q^{\alpha \beta}_{\tilde{c},\tilde{v}}  }{ \omega_{cv} -\omega - i \eta} + \text{c.c.} \right) \notag \\
&= \frac{e^3  }{ 4 \hbar \omega^2}   \sum_{\bm{k}}  d^\tau_{f} \omega^2_{cv}  \left[ \left( \text{Re}  \left( S^{\beta \gamma \alpha}_{\tilde{c},\tilde{v}}  \right) + \frac{2 \Delta^\gamma_{cv}}{\omega_{cv}} g^{\beta \alpha}_{\tilde{c},\tilde{v}}\right) \mathcal{P} (\omega_{cv} -\omega) -  \pi  \text{Im}  \left( S^{\beta \gamma \alpha}_{\tilde{c},\tilde{v}}  \right)  \delta (\omega_{cv} -\omega)   \right] ,
\label{eq:S3-19}
\end{align}
where I use the relation $(\omega_{cv} -\omega -i \eta )^{-1}= \mathcal{P} (\omega_{cv} -\omega) + i \pi \delta(\omega_{cv} -\omega) $ with $\mathcal{P}$ denoting the principal value. The principal value corresponds to off-resonance conditions, while the $\delta$-function selects the on-resonance contributions.

The CP light response is obtained by antisymmetrizing the $\alpha,\beta$:
\begin{align}
\chi^{\gamma\tau; \sigma}_{C,\text{shift}}  &= i \epsilon_{\alpha \beta \sigma}  \frac{e^3   }{ 8 \hbar \omega^2}   \sum_{\bm{k}}  d^\tau_{f} \omega^2_{cv} \left( \frac{ C^{\beta \gamma \alpha}_{\tilde{c},\tilde{v}} +  \frac{ \Delta^\gamma_{cv} }{\omega_{cv} } Q^{\beta\alpha}_{\tilde{c},\tilde{v}} - C^{\alpha \gamma \beta}_{\tilde{c},\tilde{v}} -   \frac{ \Delta^\gamma_{cv} }{\omega_{cv} } Q^{\alpha \beta}_{\tilde{c},\tilde{v}}  }{ \omega_{cv} -\omega - i \eta} - \text{c.c.} \right) \notag \\
&=-  \epsilon_{\alpha \beta \sigma}  \frac{e^3   }{ 4 \hbar \omega^2}   \sum_{\bm{k}}  d^\tau_{f} \omega^2_{cv} \left[ \left( \text{Im}  \left( A^{\beta \gamma \alpha}_{\tilde{c},\tilde{v}}  \right) - \frac{\Delta^\gamma_{cv}}{\omega_{cv}} \Omega^{\beta \alpha}_{\tilde{c},\tilde{v}} \right) \mathcal{P} (\omega_{cv} -\omega) + \pi  \text{Re}  \left( A^{\beta \gamma \alpha}_{\tilde{c},\tilde{v}} \right)  \delta (\omega_{cv} -\omega)   \right] .
\label{eq:S3-20}
\end{align}

In the massive Dirac electron with trivial band topology, the Berry curvature vanishes and the quantum metric connection lacks a symplectic form. Consequently, the linear shift current has only off-resonance contribution, while the circular shift current arises solely from on-resonance contribution. 

The symmetric and antisymmetric quantum metric connection for the Dirac Hamiltonian are given by
\begin{align}
\label{eq:S3-21}
S^{\beta \gamma \alpha}_{\tilde{c},\tilde{v}}  &= \frac{2 v_F^6}{ \omega_k^6} k^\gamma k^\beta k^\alpha  -\frac{ v_F^4}{ 2 \omega_k^4} \left(    2 \delta_{\beta \alpha} k^\gamma +   \delta_{\gamma \beta}   k^\alpha +  \delta_{\gamma \alpha} k^\beta   \right) ,\\
A^{\beta \gamma \alpha}_{\tilde{c},\tilde{v}} &=    \frac{v_F^4}{2  \omega_k^4}  \left(    \delta_{\gamma \alpha}  k^\beta - \delta_{\gamma \beta}  k^\alpha  \right) .
\label{eq:S3-22}
\end{align}

The off-resonant contribution of LP light response in the Dirac Hamiltonian reads
\begin{align}
\chi^{\gamma\tau; \alpha \beta}_{L,\text{shift}} 
&= \frac{e^3 v_F^4 }{ 2 \hbar \omega^2}   \sum_{\bm{k}}  \left( f'_c   -  f'_v \right) \left( \frac{2 v_F^4 k^\gamma   k^\tau k^\beta k^\alpha}{ \omega_k^5} - \frac{  \delta_{\gamma \beta}v_F^2  k^\alpha  k^\tau+  \delta_{\gamma \alpha} v_F^2 k^\beta  k^\tau}{ \omega_k^3}  \right) \mathcal{P} \left( 2 \omega_{k} -\omega \right)   .
\label{eq:S3-24}
\end{align}

The on-resonance contribution of CP light response in the Dirac Hamiltonian reads
\begin{align}
\chi^{\gamma \tau; \sigma}_{C,\text{shift}}  
&=  \epsilon_{\alpha \beta \sigma}  \frac{ \pi e^3 v_F^4   }{ 2 \hbar \omega^2}   \sum_{\bm{k}}  \left( f'_c   -  f'_v  \right)     \frac{    \delta_{\gamma \alpha} v_F^2 k^\beta k^\tau -\delta_{\gamma \beta} v_F^2 k^\alpha k^\tau  }{ \omega_k^3}  \delta\left(2 \omega_{k}  -\omega \right)   \notag \\
&=  \epsilon_{\alpha \beta \sigma}  \frac{ \pi e^3 v_F^4   }{ 2 \hbar \omega^2}   \sum_{\bm{k}}  \left( f'_c   -  f'_v  \right)     \frac{ \left( \delta_{\gamma \alpha} \delta_{\beta \tau}  - \delta_{\gamma \beta} \delta_{\alpha \tau} \right) v_F^2 k^2  }{3 \omega_k^3}  \delta\left(2 \omega_{k}  -\omega \right)   \notag \\
 &= \epsilon_{ \gamma \tau \sigma} | \chi_{C,\text{shift}} | ,
\label{eq:S3-24}
\end{align}
where I define the isotropic shift PDE tensor as
\begin{align}
 | \chi_{C,\text{shift}} | =  \frac{ \pi e^3 v_F^4  }{ 3 \hbar \omega^2}  \sum_{\bm{k}} \left( f'_c   -  f'_v  \right)     \frac{ \omega_k^2 - \tilde{\Delta}^2 }{ \omega_k^3}  \delta\left(2 \omega_{k}  -\omega \right)  .
\label{eq:S3-25}
\end{align}

\end{widetext}


\end{document}